\documentclass[twocolumn,pre,aps,showpacs,showkeys,amsmath]{revtex4}
\usepackage{graphicx}

\begin{document}

\title{Stochastic Spike Synchronization in A Small-World Neural Network with Spike-Timing-Dependent Plasticity}

\author{Sang-Yoon Kim}
\email{sykim@icn.re.kr}
\author{Woochang Lim}
\email{wclim@icn.re.kr}
\affiliation{Institute for Computational Neuroscience and Department of Science Education, Daegu National University of Education, Daegu 42411, Korea}

\begin{abstract}
We consider the Watts-Strogatz small-world network (SWN) consisting of subthreshold neurons which exhibit noise-induced spikings. This neuronal network has
adaptive dynamic synaptic strengths governed by the spike-timing-dependent plasticity (STDP). In previous works without STDP, stochastic spike synchronization (SSS)
between noise-induced spikings of subthreshold neurons was found to occur in a range of intermediate noise intensities. Here, we investigate the effect of additive STDP
on the SSS by varying the noise intensity. Occurrence of a ``Matthew'' effect in synaptic plasticity is found due to a positive feedback process. As a result, good synchronization
gets better via long-term potentiation of synaptic strengths, while bad synchronization gets worse via long-term depression.
Emergences of long-term potentiation and long-term depression of synaptic strengths are intensively investigated via microscopic studies based on the pair-correlations between the pre- and the post-synaptic
IISRs (instantaneous individual spike rates) as well as the distributions of time delays between the pre- and the post-synaptic spike times. Furthermore, the effects of
multiplicative STDP (which depends on states) on the SSS are studied and discussed in comparison with the case of additive STDP (independent of states).
These effects of STDP on the SSS in the SWN are also compared with those in the regular lattice and the random graph.
\end{abstract}


\maketitle

\section{Introduction}
\label{sec:INT}
In recent years, much attention has been paid to brain rhythms \cite{Buz,TW}. These brain rhythms emerge via population synchronization between individual firings in neural circuits.
This kind of neural synchronization is associated with diverse  cognitive functions (e.g., multisensory feature integration, selective attention, and memory formation) \cite{W_Review,Gray},
and it is also correlated with pathological rhythms related to  neural diseases (e.g., tremors in the Parkinson's disease and epileptic seizures) \cite{ND1,ND2}.
Population synchronization has been intensively investigated in neural circuits composed of spontaneously-firing suprathreshold neurons exhibiting regular discharges like clock
oscillators \cite{W_Review}. In contrast to the case of suprathreshold neurons, the case of subthreshold neurons (which cannot fire spontaneously) has received little attention. The
subthreshold neurons can fire only with the help of noise, and exhibit irregular discharges like Geiger counters. Noise-induced firing patterns of subthreshold neurons have been studied
in many physiological and pathophysiological aspects \cite{Braun1}. For example, sensory receptor neurons were found to use the noise-induced firings for encoding environmental electric
or thermal stimuli through a ``constructive'' interplay of subthreshold oscillations and noise \cite{Braun2}. These noise-induced firings of a single subthreshold neuron become most coherent
at an optimal noise intensity, which is called coherence resonance \cite{Longtin}. Moreover, array-enhanced coherence resonance was also found to occur in a population of subthreshold
neurons \cite{CR1,CR2,CR3,CR4,CR5}. In this way, in certain circumstances, noise plays a constructive role in the emergence of dynamical order, although it is usually considered as a nuisance,
degrading the performance of dynamical systems.

Here, we are interested in stochastic spike synchronization (SSS) (i.e., population synchronization between complex noise-induced firings of subthreshold neurons) which may
be correlated with brain function of encoding sensory stimuli in the noisy environment. Recently, such SSS has been found to occur in an intermediate range of noise intensity via competition
between the constructive and the destructive roles of noise \cite{Lim1,Lim2,Lim3}. As the noise intensity passes a lower threshold, a transition to SSS occurs because of a constructive role
of noise to stimulate coherence between noise-induced spikings. However, when passing a higher threshold, another transition from SSS to desynchronization takes place due to a destructive
role of noise to spoil the SSS. In the previous works on SSS, synaptic coupling strengths were static. However, in real brains synaptic strengths may vary to adapt to the environment
(i.e., they can be potentiated \cite{LTP1,LTP2,LTP3} or depressed \cite{LTD1,LTD2,LTD3,LTD4}). These adjustments of synapses are called the synaptic plasticity which provides the basis for
learning, memory, and development \cite{Abbott1}. Regarding the synaptic plasticity, we consider a Hebbian spike-timing-dependent plasticity (STDP) \cite{EtoE0,EtoE1,EtoE2,EtoE3,EtoE4,EtoE5,EtoE6,EtoE7,EtoE8,STDP1,STDP2,STDP3,STDP4,STDP5,STDP6,STDP7,STDP8}. For the STDP, the synaptic strengths vary via a Hebbian plasticity rule depending
on the relative time difference between the pre- and the post-synaptic spike times. When a pre-synaptic spike precedes a post-synaptic spike, long-term potentiation occurs; otherwise, long-term
depression appears. The effects of STDP on population synchronization in networks of (spontaneously-firing) suprathreshold neurons were studied in various aspects \cite{Brazil1,Brazil2,Tass1,Tass2}.

In this paper, we consider an excitatory Watts-Strogatz small-world network (SWN) of subthreshold neurons \cite{SWN1,SWN2,SWN3}, and investigate the effect of additive STDP (independent of states) 
on the SSS by varying the noise intensity $D$. A Matthew effect in synaptic plasticity is found to occur due to a positive feedback process. Good synchronization gets better via long-term potentiation of synaptic strengths, 
while bad synchronization gets worse via long-term depression. As a result, a step-like rapid transition to SSS occurs by changing $D$, in contrast to the relatively smooth transition in the absence of STDP. Emergences of 
long-term potentiation and long-term depression of synaptic strengths are intensively studied through microscopic investigations based on both the distributions of time delays between the pre- and the post-synaptic spike times and the
pair-correlations between the pre- and the post-synaptic IISRs (instantaneous individual spike rates). Moreover, the effects of multiplicative STDP (which depends on states) on the SSS
are also studied \cite{Tass1,Multi}. For the multiplicative case, a change in synaptic strengths scales linearly with the distance to the higher and the lower bounds of synaptic strengths, and
hence the bounds for the synaptic strength become ``soft,'' in contrast to the hard bounds for the additive case. The effects of STDP for the multiplicative case with soft bounds are discussed
in comparison with the additive case with hard bounds. Moreover, the effects of additive and multiplicative STDP on the SSS in the SWN are also compared with those in the regular lattice and the random graph.

This paper is organized as follows. In Sec.~\ref{sec:SWN}, we describe an excitatory Watts-Strogatz SWN of subthreshold Izhikevich regular spiking neurons \cite{Izhi1,Izhi2}, and the governing
equations for the population dynamics are given. Then, in Sec.~\ref{sec:STDP} we investigate the effects of STDP on the SSS for both the additive and the multiplicative cases by varying $D$.
Finally, in Sec.~\ref{sec:SUM} a summary is given.

\section{Excitatory Small-World Network of Subthreshold Neurons with Synaptic Plasticity}
\label{sec:SWN}
We consider an excitatory directed Watts-Strogatz SWN, composed of $N$ subthreshold regular spiking neurons equidistantly placed on a one-dimensional ring of radius $N/ 2 \pi$. The
Watts-Strogatz SWN interpolates between a regular lattice with high clustering (corresponding to the case of $p=0$) and a random graph with short average path length (corresponding to the case
of $p=1$) via random uniform rewiring with the probability $p$ \cite{SWN1,SWN2,SWN3}. For $p=0,$ we start with a directed regular ring lattice with $N$ nodes where each node is coupled to its
first $M_{syn}$ neighbors ($M_{syn}/2$ on either side) via outward synapses, and rewire each outward connection uniformly at random over the whole ring with the probability $p$ (without
self-connections and duplicate connections). This Watts-Strogatz SWN model may be regarded as a cluster-friendly extension of the random network by reconciling the six degrees of separation
(small-worldness) \cite{SDS1,SDS2} with the circle of friends (clustering). Many recent works on various subjects of neurodynamics have been done in SWNs with predominantly local
connections and rare long-distance connections \cite{SW1,SW2,SW3,SW4,SW5,SW6,SW7,SW8,SW9,SW10,SW11,SW12,SW13}. As elements in our SWN, we choose the Izhikevich regular spiking neuron model which
is not only biologically plausible, but also computationally efficient \cite{Izhi1,Izhi2}.

The following equations (\ref{eq:PD1})-(\ref{eq:PD6}) govern the population dynamics in the SWN:
\begin{eqnarray}
\frac{dv_i}{dt} &=& F(v_i) - u_i + I_{DC,i} +D \xi_{i} -I_{syn,i}, \label{eq:PD1} \\
\frac{du_i}{dt} &=& a~ (b v_i - u_i),  \;\;\; i=1, \cdots, N, \label{eq:PD2}
\end{eqnarray}
with the auxiliary after-spike resetting:
\begin{equation}
{\rm if~} v_i \geq v_p,~ {\rm then~} v_i \rightarrow c~ {\rm and~} u_i \rightarrow u_i + d, \label{eq:PD3}
\end{equation}
where
\begin{eqnarray}
F(v) &=& 0.04 v^2 + 5 v + 140,  \label{eq:PD4} \\
I_{syn,i} &=& \frac{1}{d_i^{(in)}} \sum_{j=1 (j \ne i)}^N J_{ij}~w_{ij}~s_j(t)~ (v_i - V_{syn}), \label{eq:PD5}\\
s_j(t) &=& \sum_{f=1}^{F_j} E(t-t_f^{(j)}-\tau_l); \nonumber \\
E(t) &=& \frac{1}{\tau_d - \tau_r} (e^{-t/\tau_d} - e^{-t/\tau_r}) \Theta(t). \label{eq:PD6}
\end{eqnarray}
Here, $v_i(t)$ and $u_i(t)$ are the state variables of the $i$th neuron at a time $t$ which represent the membrane potential and the recovery current, respectively. These membrane
potential and the recovery variable, $v_i(t)$ and $u_i(t)$, are reset according to Eq.~(\ref{eq:PD3}) when $v_i(t)$ reaches its cutoff value $v_p$. The parameter values used in our
computations are listed in Table \ref{tab:Parm}. More details on the Izhikevich regular spiking neuron model, the external stimulus to each Izhikevich regular spiking neuron, the synaptic currents and plasticity,
and the numerical method for integration of the governing equations are given in the following subsections.

\begin{table}
\caption{Parameter values used in our computations; units of the potential and the time are mV and msec, respectively.}
\label{tab:Parm}
\begin{ruledtabular}
\begin{tabular}{llllll}
(1) & \multicolumn{5}{l}{Single Izhikevich Regular Spiking Neurons \cite{Izhi1,Izhi2}} \\
&  $a=0.02$ & $b=0.2$ & $c=-65$ & $d=8$ & $v_p=30$ \\
\hline
(2) & \multicolumn{5}{l}{External Stimulus to Izhikevich Regular Spiking Neurons} \\
& \multicolumn{2}{l}{$I_{DC,i} \in [3.55, 3.65]$} & \multicolumn{3}{l}{$D$: Varying} \\
\hline
(3) & \multicolumn{5}{l}{Excitatory Synapse Mediated by The AMPA} \\
 & \multicolumn{5}{l}{Neurotransmitter \cite{AMPA}} \\
& $\tau_l=1$ & $\tau_r=0.5$ & $\tau_d=2$ & \multicolumn{2}{l}{$V_{syn}=0$} \\
\hline
(4) & \multicolumn{5}{l}{Synaptic Connections between Neurons in The}\\
 & \multicolumn{5}{l}{Watts-Strogatz SWN} \\
& $M_{syn}=20$ & \multicolumn{4}{l}{$p$: Varying} \\
\hline
(5) & \multicolumn{5}{l}{Hebbian STDP Rule} \\
& $A_{+} = 1.0$ & $A_{-} = 0.7$ & $\tau_{+} = 35$ & $\tau_{-} = 70$ & \\
& $\delta = 0.005$ & \multicolumn{4}{l}{$J_{ij} \in [0.0001, 1.0]$} \\
\end{tabular}
\end{ruledtabular}
\end{table}

\subsection{Izhikevich Regular Spiking Neuron Model}
\label{subsec:Izhi}
We first note that the function $F(v)$ in Eq.~(\ref{eq:PD4}) for the dynamics of the Izhikevich neuron was obtained by fitting the spike initiation dynamics of a cortical neuron so that the membrane potential 
$v$ has mV scale and the time $t$ has msec scale  \cite{Izhi1,Izhi2}. Then, the Izhikevich model may match neuronal dynamics by tuning the parameters $(a, b, c, d)$ instead of matching neuronal electrophysiology, unlike the Hodgkin-Huxley-type conductance-based models \cite{Izhi1,Izhi2}. The parameters $a$, $b$, $c$, and $d$ are related to the time scale of the recovery variable $u$, the sensitivity of $u$ to the subthreshold fluctuations of $v$, and the after-spike reset values of $v$ and $u$, respectively. Depending on the values of these parameters, the Izhikevich neuron model may exhibit 20 of the most prominent neuro-computational features of cortical
neurons \cite{Izhi1,Izhi2}. Here, we use the parameter values for the regular spiking  neurons, which are listed in the 1st item of Table \ref{tab:Parm}.

\subsection{External Stimulus to Each Izhikevich Regular Spiking Neuron}
\label{subsec:Sti}
Each Izhikevich regular spiking neuron is stimulated by both a common DC current $I_{DC,i}$ and an independent Gaussian white noise $\xi_i$ [see the 3rd and the 4th terms in Eq.~(\ref{eq:PD1})].
The Gaussian white noise satisfies $\langle \xi_i(t) \rangle =0$ and $\langle \xi_i(t)~\xi_j(t') \rangle = \delta_{ij}~\delta(t-t')$, where $\langle\cdots\rangle$ denotes an ensemble
average. Here, the intensity of the Gaussian noise $\xi$ is controlled by the parameter $D$. For $D=0$, the Izhikevich regular spiking neurons exhibit the type-II excitability. A type-II neuron
exhibits a jump from a resting state to a spiking state through a subcritical Hopf bifurcation when passing a threshold by absorbing an unstable limit cycle born via fold limit cycle
bifurcation and hence, the firing frequency begins from a non-zero value \cite{Ex1,Ex2}. Throughout the paper, we consider a subthreshold case (where only noise-induced firings occur)
such that the value of $I_{DC,i}$ is chosen via uniform random sampling in the range of [3.55, 3.65], as shown in the 2nd item of Table \ref{tab:Parm}.

\subsection{Synaptic Currents and Plasticity}
\label{subsec:Syn}
The 5th term in Eq.~(\ref{eq:PD1}) denotes the synaptic couplings of Izhikevich regular spiking neurons. $I_{syn,i}$ of Eq.~(\ref{eq:PD5}) represents the synaptic current injected into the $i$th neuron,
and $V_{syn}$ is the synaptic reversal potential. The synaptic connectivity is given by the connection weight matrix $W$ (=$\{ w_{ij} \}$) where $w_{ij}=1$ if the neuron $j$ is presynaptic
to the neuron $i$; otherwise, $w_{ij}=0$.
Here, the synaptic connection is modeled in terms of the Watts-Strogatz SWN. The in-degree of the $i$th neuron, $d_{i}^{(in)}$ (i.e., the number of synaptic inputs to the neuron $i$) is
given by $d_{i}^{(in)} =
\sum_{j=1(\ne i)}^N w_{ij}$. For this case, the average number of synaptic inputs per neuron is given by $M_{syn} = \frac{1}{N} \sum_{i=1}^{N} d_{i}^{(in)}$. Throughout the paper, $M_{syn}=20$ (see the 4th item of
Table \ref{tab:Parm}).

The fraction of open synaptic ion channels at time $t$ is denoted by $s(t)$. The time course of $s_j(t)$ of the $j$th neuron is given by a sum of delayed double-exponential functions
$E(t-t_f^{(j)}-\tau_l)$ [see Eq.~(\ref{eq:PD6})], where $\tau_l$ is the synaptic delay, and $t_f^{(j)}$ and $F_j$ are the $f$th spiking time and the total number of spikes of the $j$th neuron
(which occur until time $t$), respectively. Here, $E(t)$ [which corresponds to contribution of a pre-synaptic spike occurring at time $0$ to $s(t)$ in the absence of synaptic delay] is controlled
by the two synaptic time constants: synaptic rise time $\tau_r$ and decay time $\tau_d$, and $\Theta(t)$ is the Heaviside step function: $\Theta(t)=1$ for $t \geq 0$ and 0 for $t <0$. For the
excitatory AMPA synapse, the values of $\tau_l$, $\tau_r$, $\tau_d$, and $V_{syn}$ are listed in the 3rd item of Table \ref{tab:Parm} \cite{AMPA}.

The coupling strength of the synapse from the $j$th pre-synaptic neuron to the $i$th post-synaptic neuron is $J_{ij}$. Here, we consider a Hebbian STDP for the synaptic strengths $\{ J_{ij} \}$.
Initial synaptic strengths are normally distributed with the mean $J_0(=0.2)$ and the standard deviation $\sigma_0(=0.02)$. With increasing time $t$, the synaptic strength for each synapse is updated
with an additive nearest-spike pair-based STDP rule \cite{SS}:
\begin{equation}
  J_{ij} \rightarrow J_{ij} + \delta~ \Delta J_{ij}(\Delta t_{ij}),
\label{eq:ASTDP}
\end{equation}
where $\delta$ $(=0.005)$ is the update rate and $\Delta J_{ij}$ is the synaptic modification depending on the relative time difference $\Delta t_{ij}$ $(=t_i^{(post)} - t_j^{(pre)})$
between the nearest spike times of the post-synaptic neuron $i$ and the pre-synaptic neuron $j$.
To avoid unbounded growth and negative conductances (i.e. negative coupling strength), we set a range with the upper and the lower bounds: $J_{ij} \in [J_l, J_h]$.
Specifically, the upper bound of $J_{ij}$ is set to $J_h = 1$ to avoid occurrence of noise-induced burtsings for a strong excitatory coupling \cite{NB}, and the lower boundary is set 
to $J_l=0.0001$ (i.e., slightly greater than 0) to avoid elimination of synaptic connections.
We use an asymmetric time window for the synaptic modification $\Delta J_{ij}(\Delta t_{ij})$ \cite{STDP1}:
\begin{equation}
  \Delta J_{ij} = \left\{ \begin{array}{l} A_{+}~  e^{-\Delta t_{ij} / \tau_{+}} ~{\rm for}~ \Delta t_{ij} > 0\\
  - A_{-}~ e^{\Delta t_{ij} / \tau_{-}} ~{\rm for}~ \Delta t_{ij} < 0\end{array} \right. ,
\label{eq:TW}
\end{equation}
where $A_+=1.0$, $A_-=0.7$, $\tau_+=35$ msec, $\tau_-=70$ msec (these values are also given in the 5th item of Table \ref{tab:Parm}), and $\Delta J_{ij}(\Delta t_{ij}=0) = 0$.

\subsection{Numerical Method for Integration}
\label{subsec:NM}
Numerical integration of stochastic differential Eqs.~(\ref{eq:PD1})-(\ref{eq:PD6}) with a Hebbian STDP rule of Eqs.~(\ref{eq:ASTDP}) and (\ref{eq:TW})
is done by employing the Heun method \cite{SDE} with the time step $\Delta t=0.01$ msec. For each
realization of the stochastic process, we choose  random initial points $[v_i(0),u_i(0)]$ for the $i$th $(i=1,\dots, N)$ regular spiking neuron with uniform probability in the range of
$v_i(0) \in (-50,-45)$ and $u_i(0) \in (10,15)$.

\begin{figure}
\includegraphics[width=0.8\columnwidth]{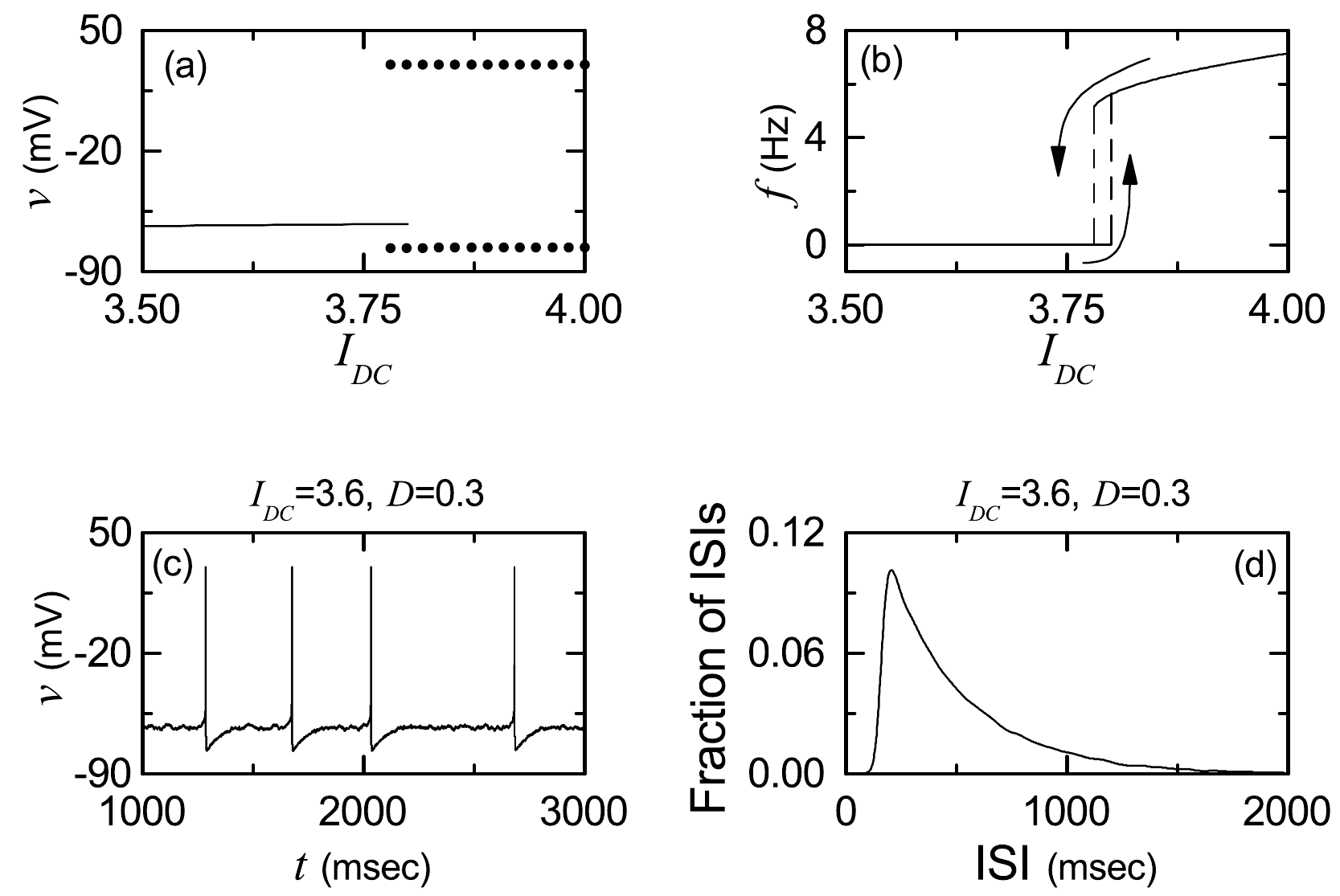}
\caption{Single Izhikevich regular spiking neuron exhibiting type-II excitability. (a) Bifurcation diagram (i.e., $v$ versus $I_{DC})$ for $D = 0$. Solid line represents a stable equilibrium point, while maximum and minimum values of $v$ for the spiking state are denoted by solid circles. (b) Plot of the mean firing rate $f$ versus $I_{DC}$ for $D = 0$. (c) Time series of $v$ for $I_{DC}$ = 3.6 and $D = 0.3$. (d) Interspike interval (ISI) histogram for $I_{DC} = 3.6$ and $D = 0.3$.
}
\label{fig:SN}
\end{figure}

\section{Effects of the STDP on the Stochastic Spike Synchronization}
\label{sec:STDP}
\begin{figure*}
\includegraphics[width=1.5\columnwidth]{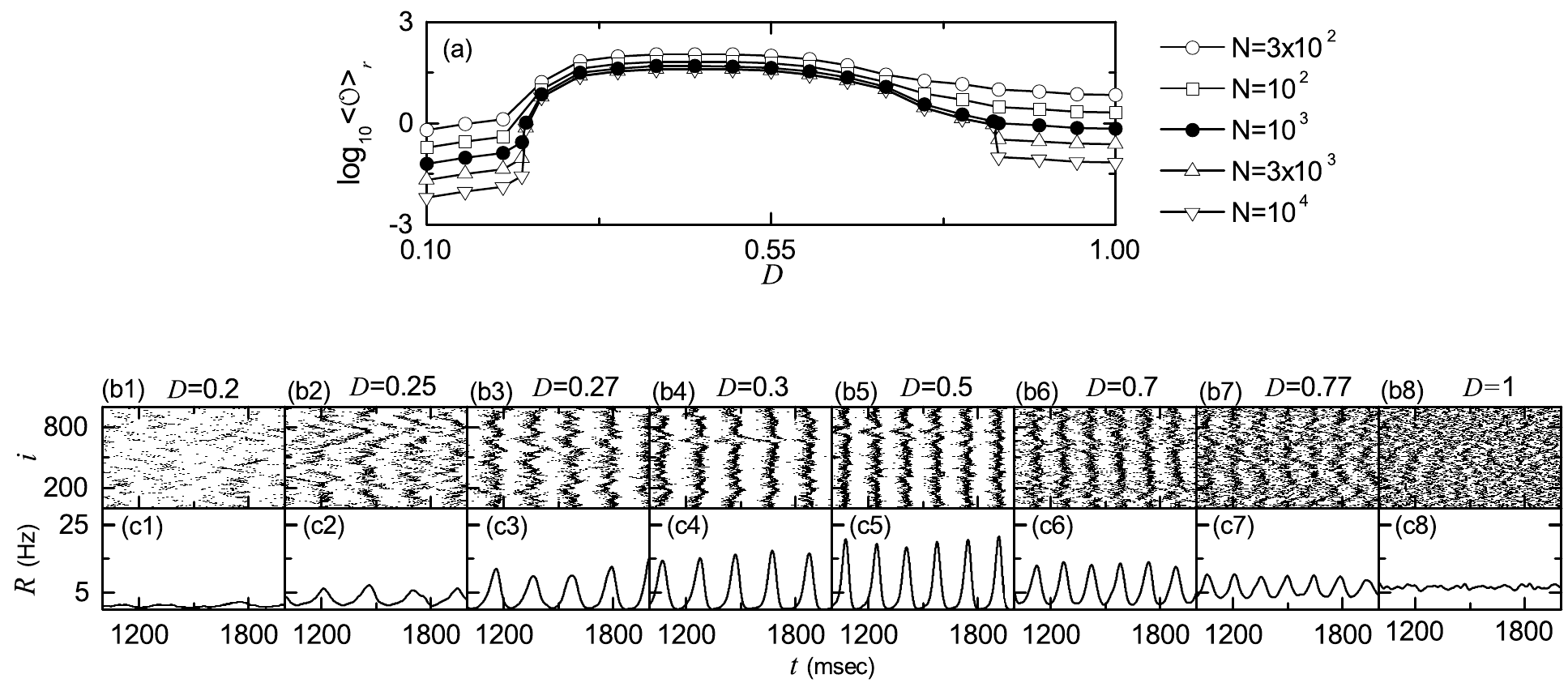}
\caption{SSS for $p=0.15$ in the absence of STDP. (a) Plots of the thermodynamic order parameter $\langle {\cal{O}} \rangle_r$  versus $D$. Raster plots of spikes in (b1)-(b8) and IPSR kernel estimates
$R(t)$ in (c1)-(c8) for various values of $D$ = 0.2, 0.25, 0.27, 0.3, 0.5, 0.7, 0.77, and 1.
}
\label{fig:NSTDP1}
\end{figure*}

We consider the Watts-Strogatz SWN with high clustering and short path length when the rewiring probability $p$ is 0.15. This SWN is composed of $N$ excitatory subthreshold
Izhikevich regular spiking neurons (exhibiting noise-induced spikings). Throughout the paper, $N=10^3$ except for the case of the order parameter in Fig.~\ref{fig:NSTDP1}(a).
As shown in Fig.~\ref{fig:SN}(a), the Izhikevich  regular spiking neuron exhibits a jump from a resting state (denoted by a solid line)
to a spiking state (represented by solid circles) via subcritical Hopf bifurcation at a higher threshold $I_{DC,h} (\simeq 3.80)$ by absorbing an unstable limit cycle born through
a fold limit cycle bifurcation for a lower threshold $I_{DC,l} (\simeq 3.78)$. Hence, the Izhikevich regular spiking neuron exhibits type-II excitability because it begins to fire with a non-zero
frequency \cite{Ex1,Ex2}. Figure \ref{fig:SN}(b) shows a plot of the mean firing rate (MFR) $f$ versus the external DC current $I_{DC}$ for a single Izhikevich regular spiking neuron in the absence
of noise ($D=0$).  As $I_{DC}$ is increased from $I_{DC,h}$, the MFR $f$ increases monotonically. As an example, we consider a subthreshold case of $I_{DC}=3.6$ in the presence of
noise with $D=0.3$ for which a time series of the membrane potential $v$ with the MFR $f \simeq 1.98$ Hz is shown in Fig.~\ref{fig:SN}(c). Figure \ref{fig:SN}(d) also shows a
histogram for distribution of interspike intervals (ISIs) for $I_{DC}=3.6$ and $D=0.3$. The average ISI $\langle ISI \rangle$ is 506.3 msec; the reciprocal of $\langle ISI \rangle$
corresponds to the MFR. This distribution is also broad because of a large standard deviation (=350.2 msec) from the average value.

\subsection{SSS in The Absence of STDP}
\label{subsec:NSTDP}
First, we are concerned about the SSS in the absence of STDP. The coupling strengths $\{ J_{ij} \}$ are static, and their values are chosen from the Gaussian distribution where the mean
$J_0$ is 0.2 and the standard deviation $\sigma_0$ is 0.02. Population synchronization may be well visualized in the raster plot of neural spikes which is a collection of spike
trains of individual neurons. Such raster plots of spikes are fundamental data in experimental neuroscience. As a collective quantity showing population behaviors,
we use an instantaneous population spike rate (IPSR) which may be obtained from the raster plots of spikes \cite{W_Review,RM}. For the synchronous case, ``stripes" (composed of spikes
and indicating population synchronization) are found to be formed in the raster plot, while in the unsynchronized case spikes are completely scattered. Hence, for a synchronous case,
an oscillating IPSR $R(t)$ appears, while for an unsynchronized case $R(t)$ is nearly stationary. To obtain a smooth IPSR, we employ the kernel density estimation (kernel smoother)
\cite{Kernel}. Each spike in the raster plot is convoluted (or blurred) with a kernel function $K_h(t)$ to obtain a smooth estimate of IPSR $R(t)$:
\begin{equation}
R(t) = \frac{1}{N} \sum_{i=1}^{N} \sum_{s=1}^{n_i} K_h (t-t_{s}^{(i)}),
\label{eq:IPSR}
\end{equation}
where $t_{s}^{(i)}$ is the $s$th spiking time of the $i$th neuron, $n_i$ is the total number of spikes for the $i$th neuron, and we use a Gaussian
kernel function of band width $h$:
\begin{equation}
K_h (t) = \frac{1}{\sqrt{2\pi}h} e^{-t^2 / 2h^2}, ~~~~ -\infty < t < \infty.
\label{eq:Gaussian}
\end{equation}
Throughout the paper, the band width $h$ of $K_h(t)$ is 10 msec. Recently, we introduced a realistic thermodynamic order parameter, based on $R(t)$, for describing transition from
desynchronization to synchronization \cite{RM}.
The mean square deviation of $R(t)$,
\begin{equation}
{\cal{O}} \equiv \overline{(R(t) - \overline{R(t)})^2},
 \label{eq:Order}
\end{equation}
plays the role of an order parameter $\cal{O}$; the overbar represents the time average. This order parameter may be regarded as a thermodynamic measure because it
concerns just the macroscopic IPSR kernel estimate $R(t)$ without any consideration between $R(t)$ and microscopic individual spikes. In the thermodynamic limit of
$N \rightarrow \infty$, the order parameter $\cal{O}$ approaches a non-zero (zero) limit value for the synchronized (unsynchronized) state. 
Figure \ref{fig:NSTDP1}(a)shows plots of $\log_{10} \langle {\cal O} \rangle_r$ versus $D$ in the SWN with for $p=0.15$. 
In each realization, we discard the first time steps of a stochastic trajectory as transients for $10^3$ msec, and then we numerically compute $\cal{O}$ by following the stochastic 
trajectory for $3 \times 10^4$ msec. Throughout the paper, $\langle \cdots \rangle_r$ denotes an average over 20 realizations. With increasing $N$ up to $10^4$, these numerical 
calculations for $\langle {\cal{O}} \rangle_r$ are done for various values of $D$. For $D < D^*_l$ $(\simeq 0.225$), unsynchronized states exist because the order parameter 
$\langle {\cal{O}} \rangle_r$ tends to decrease to zero as $N$ is increased. As $D$ passes the lower threshold $D^*_l$, $\langle {\cal{O}} \rangle_r$ tends to converge toward non-zero 
limit values, and hence a transition to SSS occurs thanks to a constructive role of noise to stimulate coherence between noise-induced spikings of subthreshold neurons. 
However, for large $D > D^*_h$ $(\simeq 0.846)$, with increasing $N$ the order parameter $\langle {\cal{O}} \rangle_r$ tends to approach zero, and hence
SSS disappears (i.e., a transition to desynchronization occurs when $D$ passes the higher threshold $D^*_h$) due to a destructive role of noise to spoil the SSS. 
In this way, SSS appears in an intermediate range of $D^*_l < D < D^*_h$ via competition between the constructive and the destructive roles of noise.
Figures \ref{fig:NSTDP1}(b1)-\ref{fig:NSTDP1}(b8) show raster plots of spikes for various values of $D$, and their corresponding IPSR kernel estimates $R(t)$ are also shown in
Figs.~\ref{fig:NSTDP1}(c1)-\ref{fig:NSTDP1}(c8). For $D=0.2$ (less than $D^*_l$), spikes are scattered without forming any stripes in the raster plot, and hence
the IPSR kernel estimate $R(t)$ is nearly stationary. On the other hand, when passing $D^*_l$, synchronized states appear. For $D=0.25$ the raster plot of spikes shows a zigzag
pattern intermingled with inclined partial stripes of spikes due to local clustering, and the IPSR kernel estimate $R(t)$ exhibits an oscillatory behavior. With increasing
$D$ the degree of SSS is increased because clearer stripes with reduced zigzagness appear (e.g., see the cases of $D=0.27,$ 0.3, and 0.5). As a result, the amplitude of
$R(t)$ increases with $D$. However, with further increase in $D$, stripes are smeared, as shown in the cases of $D=0.7$ and 0.77, and hence the amplitude of $R(t)$ decreases.
Eventually, when passing $D^*_h$ desynchronization occurs due to overlap of smeared stripes (e.g., see the case of $D=1$). We also note that the population frequency
$f_p$ of the IPSR kernel estimate $R(t)$ increases with $D$ in the range of SSS (i.e., the interval between stripes in the raster plots of spikes decreases with $D$).

\begin{figure}
\includegraphics[width=0.7\columnwidth]{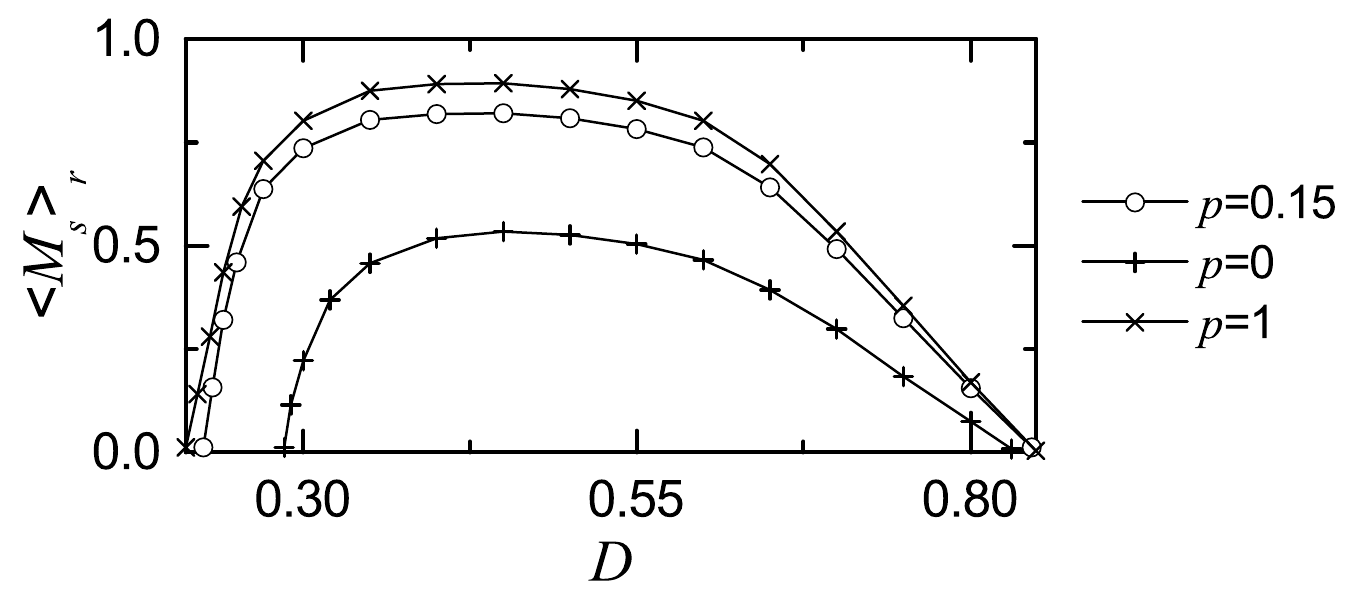}
\caption{Characterization of SSS for various values of $p$ in the absence of STDP. Plots of the statistical-mechanical spiking measure $\langle M_s \rangle_r$ versus $D$ for $p=$ 0 (pluses), 0.15 (circles),
and 1 (crosses).
}
\label{fig:NSTDP2}
\end{figure}

We characterize the SSS by employing the statistical-mechanical spiking measure $M_s$ \cite{RM}. For the case of SSS, stripes appear regularly in the raster plot of spikes.
The spiking measure $M_i$ of the $i$th stripe is defined by the product of the occupation degree $O_i$ of spikes (representing the density of the $i$th stripe) and the
pacing degree $P_i$ of spikes (denoting the smearing of the $i$th stripe):
\begin{equation}
M_i = O_i \cdot P_i.
\label{eq:SMi}
\end{equation}
The occupation degree $O_i$ of spikes in the stripe is given by the fraction of spiking neurons:
\begin{equation}
   O_i = \frac {N_i^{(s)}} {N},
\end{equation}
where $N_i^{(s)}$ is the number of spiking neurons in the $i$th stripe. For the full occupation $O_i=1$, while for the partial occupation $O_i<1$.
In our case of SSS, $O_i=1$, independently of $D$. For this case of full synchronization, $M_i = P_i$. The pacing degree $P_i$ of spikes in the $i$th stripe
can be determined in a statistical-mechanical way by taking into account their contributions to the macroscopic IPSR kernel estimate $R(t)$.
Central maxima of $R(t)$ between neighboring left and right minima of $R(t)$ coincide with centers of stripes in the raster plot. A global cycle starts from a left minimum of
$R(t)$, passes a maximum, and ends at a right minimum. An instantaneous global phase $\Phi(t)$ of $R(t)$ was introduced via linear interpolation in the region forming a global cycle
(for details, refer to Eqs.~(16) and (17) in \cite{RM}).  Then, the contribution of the $k$th microscopic spike in the $i$th stripe occurring at the time $t_k^{(s)}$ to $R(t)$ is
given by $\cos \Phi_k$, where $\Phi_k$ is the global phase at the $k$th spiking time [i.e., $\Phi_k \equiv \Phi(t_k^{(s)})$]. A microscopic spike makes the most constructive (in-phase)
contribution to $R(t)$ when the corresponding global phase $\Phi_k$ is $2 \pi n$ ($n=0,1,2, \dots$), while it makes the most destructive (anti-phase) contribution to $R(t)$ when $\Phi_k$
is $2 \pi (n-1/2)$. By averaging the contributions of all microscopic spikes in the $i$th stripe to $R(t)$, we obtain the pacing degree of spikes in the $i$th stripe:
\begin{equation}
 P_i ={ \frac {1} {S_i}} \sum_{k=1}^{S_i} \cos \Phi_k,
\label{eq:PACING}
\end{equation}
where $S_i$ is the total number of microscopic spikes in the $i$th stripe.
By averaging $P_i$ over a sufficiently large number $N_s$ of stripes, we obtain the realistic statistical-mechanical spiking measure $M_s$, based on the IPSR kernel estimate $R(t)$:
\begin{equation}
M_s =  {\frac {1} {N_s}} \sum_{i=1}^{N_s} P_i.
\label{eq:SM}
\end{equation}
\begin{figure}
\includegraphics[width=0.7\columnwidth]{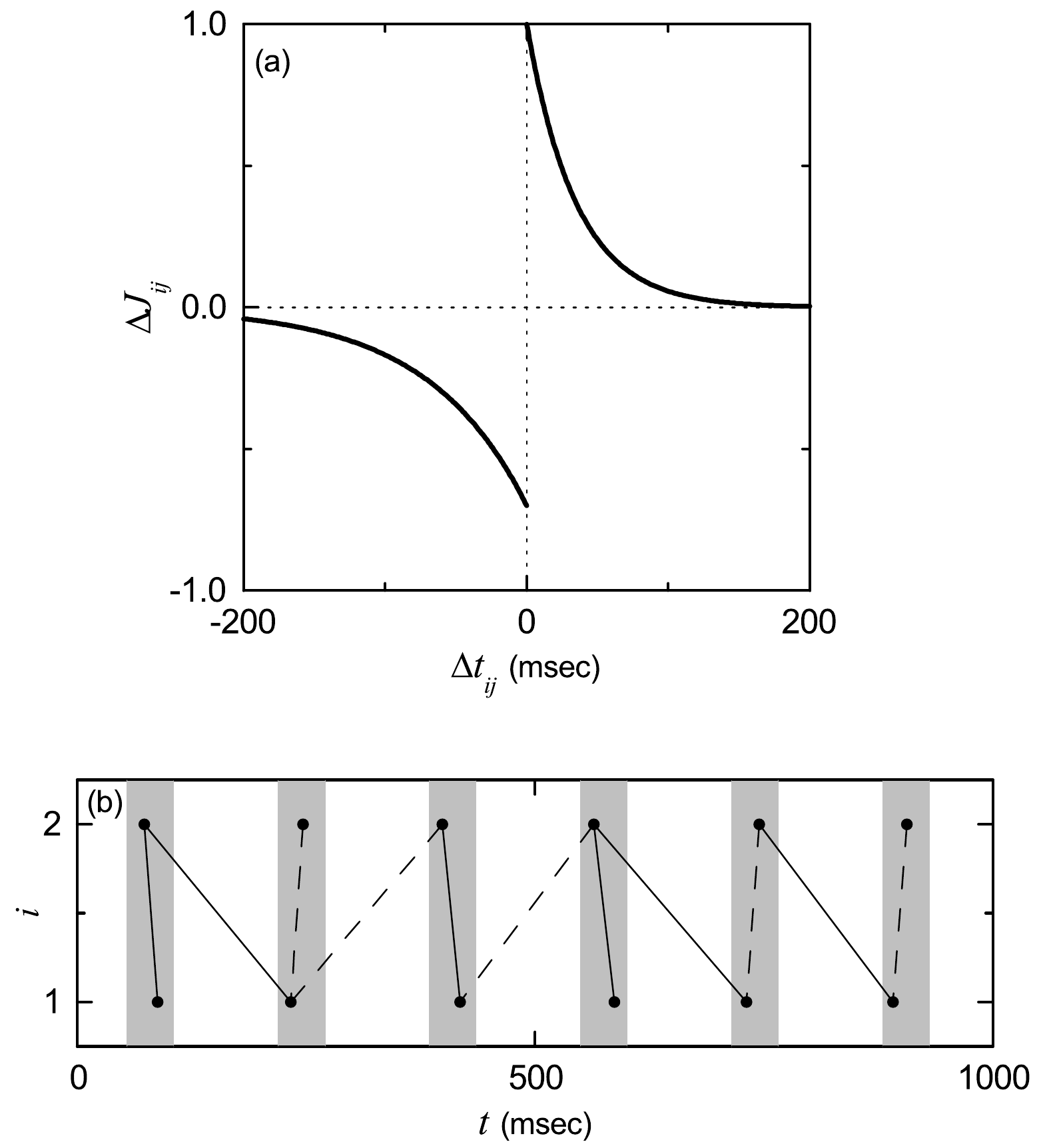}
\caption{(a) Time window for the Hebbian STDP. Plot of synaptic modification $\Delta J_{ij}$ versus $\Delta t_{ij}$ $(=t_i^{(post)} - t_j^{(pre)})$ for $A_+=1$, $A_{-}=0.7$, $\tau_+=35$ msec and $\tau_{-}=70$ msec. $t_i^{(post)}$ and $t_j^{(pre)}$ are spiking times of the $i$th post-synaptic and the $j$th pre-synaptic neurons, respectively. (b) Schematic diagram for the nearest-spike pair-based STDP rule; $i=1$ and 2 correspond to the post- and the pre-synaptic neurons. Gray boxes and solid circles denote stripes and spikes, respectively. Solid and dashed lines denote long-term potentiation and long-term depression, respectively.
}
\label{fig:TW}
\end{figure}
We follow $3 \times 10^3$ stripes in each realization and get $\langle M_s \rangle_r$ via average over 20 realizations. Figure \ref{fig:NSTDP2} shows a plot of $\langle M_s \rangle_r$
(denoted by open circles) versus $D$ in the SWN with $p=0.15$. When passing $D^*_l$ a rapid increase in $\langle M_s \rangle_r$ occurs, then a flat ``plateau'' of $\langle M_s \rangle_r$ appears, and finally
$\langle M_s \rangle_r$ decreases in a relatively slow way. Thus, a bell-shaped curve (composed of open circles) is formed.
For comparison, we also consider the cases of $p=0$ (regular lattice) and $p=1$ (random graph); the cases of $p=0$ and 1 are represented by pluses and crosses, respectively.
The topological properties of the small-world connectivity has been well characterized in terms of the clustering coefficient $C$ and the average path length $L$ \cite{SWN1}.
The clustering coefficient $C$, representing the cliquishness of a typical neighborhood in the network, characterizes the local efficiency of information transfer, while the average
path length $L$, denoting the typical separation between two vertices in the network, characterizes the global efficiency of information transfer. Particularly, short path length may
be efficient for global communication between distant neurons (i.e. neural synchronization). The regular lattice for $p=0$ is highly clustered but large world where the average path
length grows linearly with $N$ \cite{SWN1}; $C \simeq 0.71$ and $L \simeq 25.5$ for $N=10^3$. On the other hand, the random graph for $p=1$ is poorly clustered but small world where
the average path length grows logarithmically with $N$ \cite{SWN1}; $C \simeq 0.02$ and $L \simeq 2.64$ for $N=10^3$. As soon as $p$ increases from zero, the average path length $L$
decreases dramatically, which leads to occurrence of a small-world phenomenon which is popularized by the phrase of the ``six degrees of separation'' \cite{SDS1,SDS2}. However,
during such dramatic drop in $L$, the clustering coefficient $C$ decreases only a little. Consequently, for small $p$ (=0.15) an SWN with short path length $L (\simeq 3.04)$ and
high clustering $C (\simeq 0.45)$ emerges. $L$ for $p=0.15$ is much smaller than that for $p=0$ (regular lattice), and it is just a little larger than that for $p=1$ (random graph).
Hence, the values of $\langle M_s \rangle_r$ for $p=0.15$ are much larger than those for $p=0$, and they are somewhat close to those for $p=1$. However, unlike the case of $p=1$,
zigzag patterns of partially inclined stripes appear in the raster plot of spikes for $p=0.15$ due to high local clustering, as shown in Figs.~\ref{fig:NSTDP1}(b2)-\ref{fig:NSTDP1}(b7).

\subsection{Effects of The Additive STDP on The SSS}
\label{subsec:ASTDP}
From now on, we study the effect of additive STDP on the SSS. The initial values of synaptic strengths $\{ J_{ij} \}$ are chosen from the Gaussian distribution where the mean $J_0$ is 0.2 and the standard
deviation $\sigma_0$ is 0.02. Then, $J_{ij}$ for each synapse is updated according to the additive nearest-spike pair-based STDP rule of Eq.~(\ref{eq:ASTDP}) \cite{SS}. Figure \ref{fig:TW}(a) shows the time
window for the synaptic modification $\Delta J_{ij}$ of Eq.~(\ref{eq:TW}) (i.e., plot of $\Delta J_{ij}$ versus $\Delta t_{ij}$). $\Delta J_{ij}$ varies depending on the relative time difference
$\Delta t_{ij}$ $(=t_i^{(post)} - t_j^{(pre)})$ between the nearest spike times of the post-synaptic neuron $i$ and the pre-synaptic neuron $j$. When a post-synaptic spike follows a
pre-synaptic spike (i.e., $\Delta t_{ij}$ is positive), long-term potentiation of synaptic strength appears; otherwise (i.e., $\Delta t_{ij}$ is negative), long-term depression occurs. A schematic diagram for the
nearest-spike pair-based STDP rule is given in Fig.~\ref{fig:TW}(b), where $i=1$ and 2 correspond to the post- and the pre-synaptic neurons. Here, gray boxes represent stripes in
the raster plot, and spikes in the stripes are denoted by solid circles. When the post-synaptic neuron ($i=1$) fires a spike, long-term potentiation (denoted by solid lines) occurs via STDP between the
post-synaptic spike and the previous nearest pre-synaptic spike. In contrast, when the pre-synaptic neuron ($i=2$) fires a spike, long-term depression (represented by dashed lines) occurs through STDP
between the pre-synaptic spike and the previous nearest post-synaptic spike. We note that such long-term potentiation and long-term depression may occur between the pre- and the post-synaptic spikes in the same stripe or
in the different nearest-neighboring stripes; solid/dashed lines connect pre- and post-synaptic spikes in the same stripe or in the different nearest-neighboring stripes.

\begin{figure}
\includegraphics[width=\columnwidth]{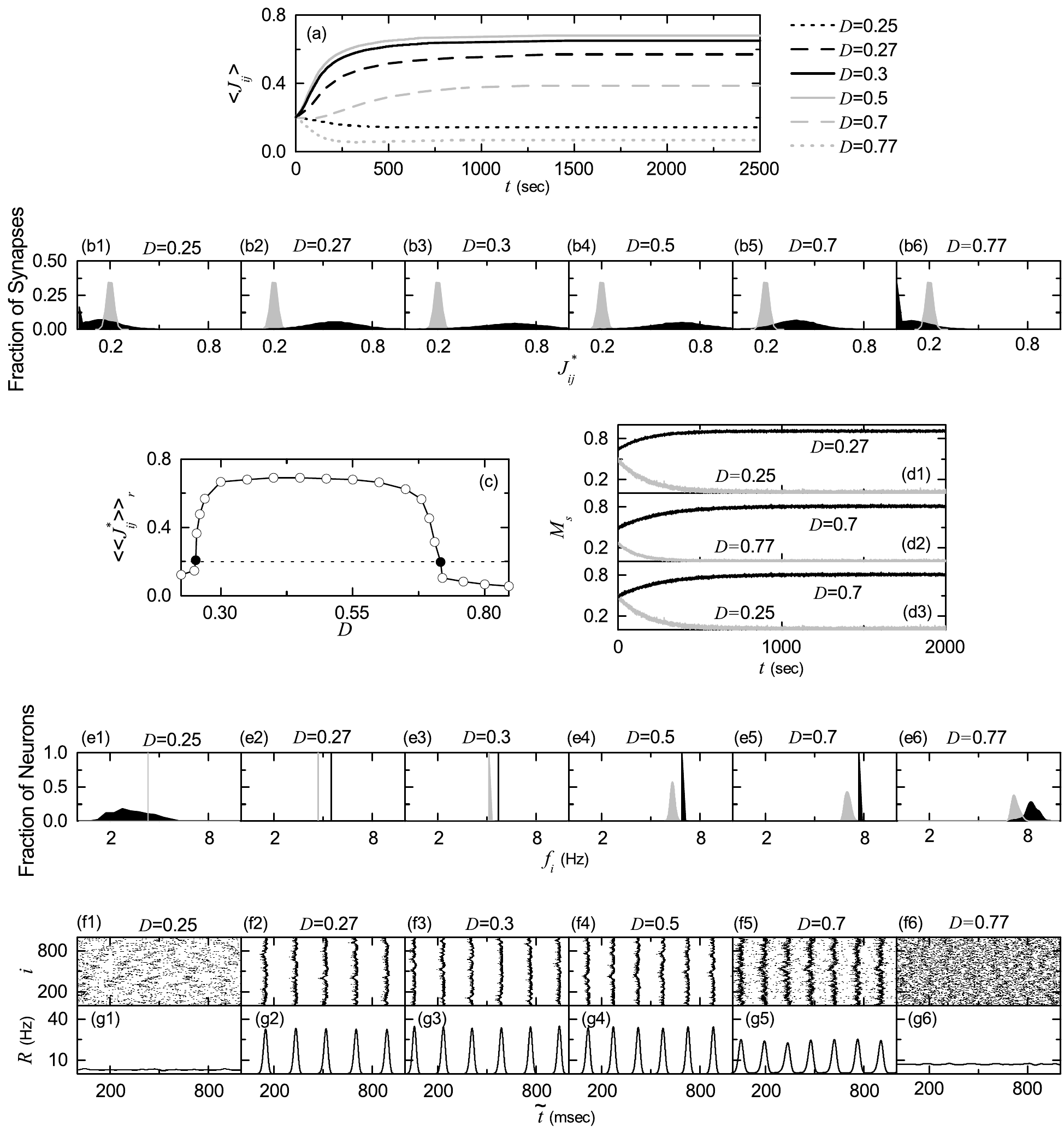}
\caption{Effects of the additive STDP on the SSS in the SWN with $p=0.15$. (a) Time-evolutions of population-averaged synaptic strengths $\langle J_{ij} \rangle$ for various values of $D$.
(b1)-(b6) Histograms for the fraction of synapses versus $J^*_{ij}$ (saturated limit values of $J_{ij}$ at $t = 2000$ sec) are shown in black color for various values of $D$;
for comparison, initial distributions of synaptic strengths $\{ J_{ij} \}$ are also shown in gray color. (c) Plot of population-averaged limit values of synaptic strengths
$\langle \langle J^*_{ij} \rangle \rangle_r$  versus $D$. Time-evolutions of statistical-mechanical spiking measure $M_s$ for (d1) $D = 0.25$ and 0.27, (d2) $D = 0.77$ and 0.7, and (d3)
$D = 0.25$ and 0.7. (e1)-(e6) Histograms for the MFRs $f_i$ of individual neurons are shown in black color for various values of $D$; for comparison, initial distributions of $\{ f_i \}$
are also shown in gray color. Raster plots of spikes in (f1)-(f6) and IPSR kernel estimates $R(t)$ in (g1)-(g6) for various values of $D$ after the saturation time, where
$t=t^*$ (saturation time=2000 sec) + $\widetilde{t}$.
}
\label{fig:STDP1}
\end{figure}

Figure \ref{fig:STDP1}(a) shows time-evolutions of population-averaged synaptic strengths $\langle J_{ij} \rangle$ for various values of $D$ in the SWN with $p=0.15$; $\langle \cdots \rangle$ represents an
average over all synapses. For each case of $D=0.27,$ 0.3, 0.5 and 0.7, $\langle J_{ij} \rangle$ increases monotonically above its initial value $J_0$ (=0.2), and it approaches a saturated limit value $\langle
J_{ij}^* \rangle$ nearly at $t=2000$ sec. Consequently, long-term potentiation occurs for these values of $D$. On the other hand, for $D=0.25$ and 0.77, $\langle J_{ij} \rangle$ decreases monotonically below $J_0$, and
approaches a saturated limit value $\langle J_{ij}^* \rangle$. As a result, long-term depression occurs for the cases of $D=0.25$ and 0.77. Histograms for fraction of synapses versus $J_{ij}^*$ (saturated limit values
of $J_{ij}$ at $t=2000$ sec) are shown in black color for various values of $D$ in Figs.~\ref{fig:STDP1}(b1)-\ref{fig:STDP1}(b6); the bin size for each histogram is 0.02. For comparison,
initial distributions of synaptic strengths $\{ J_{ij} \}$ (i.e., Gaussian distributions whose mean $J_0$ and standard deviation $\sigma_0$ are 0.2 and 0.02, respectively) are also shown in gray color. For the
cases of long-term potentiation ($D=0.27,$ 0.3, 0.5 and 0.7), their black histograms lie on the right side of the initial gray histograms, and hence their population-averaged values $\langle J_{ij}^*
\rangle$ become larger than the initial value $J_0$ (=0.2). In contrast, the black histograms for the cases of long-term depression ($D=0.25$ and 0.77) are shifted to the left side of the initial gray
histograms, and hence their population-averaged values $\langle J_{ij}^* \rangle$ become smaller than $J_0$. For both cases of long-term potentiation and long-term depression, their black histograms are much wider than the initial
gray histograms [i.e., the standard deviations $\sigma$ are very larger than the initial one $\sigma_0$ (=0.02)]. Figure \ref{fig:STDP1}(c) shows a plot of population-averaged limit values of synaptic strengths
$\langle \langle J_{ij}^* \rangle \rangle_r$ versus $D$. Here, the horizontal dotted line represents the initial average value of coupling strengths $J_0$ (= 0.2), and the lower and the higher threshold
values $\widetilde{D}_l$ $(\simeq 0.253)$ and $\widetilde{D}_h$ $(\simeq 0.717)$ for long-term potentiation and long-term depression (where $\langle \langle J_{ij}^* \rangle \rangle_r = J_0$) are denoted by solid circles. Hence, long-term potentiation occurs in the range of ($\widetilde{D}_l$, $\widetilde{D}_h$); otherwise, long-term depression appears. We also note that the range of ($\widetilde{D}_l$, $\widetilde{D}_h$) is strictly contained in the range of ($D^*_l$, $D^*_h$)
($D^*_l \simeq 0.225$ and $D^*_h \simeq 0.846$) where SSS appears in the absence of STDP [i.e., ($\widetilde{D}_l$, $\widetilde{D}_h$) is a proper subset of ($D^*_l$, $D^*_h$)].
Hence, in most range of the SSS long-term potentiation occurs, while long-term depression takes place only near both ends. Similar to the case in Fig.~\ref{fig:STDP1}(c), a bell-shaped curve (showing a plot of average synaptic strengths
versus noise intensity) was also observed for the case where many nearly coincident pre-synaptic inputs are given to a post-synaptic neuron \cite{Aihara}.

We now consider the effects of long-term potentiation and long-term depression on the SSS for $p=0.15$. Time-evolutions of the statistical-mechanical spiking measures $M_s$ [of Eq.~(\ref{eq:SM})] for the population states are shown
in Figs.~\ref{fig:STDP1}(d1)-\ref{fig:STDP1}(d3); black (gray) curves represent the cases of long-term potentiation (long-term depression). For the case of close small values of $D$ in Fig.~\ref{fig:STDP1}(d1), the initial value
of $M_s$ for $D=0.27$ is a little larger than that for $D=0.25$. However, with increasing time $t$, $M_s$ for $D=0.27$ increases thanks to long-term potentiation, and it approaches its limit value. On the other hand,
$M_s$ for $D=0.25$ decreases due to long-term depression, and it seems to approach zero (i.e., desynchronization occurs). A similar one takes place for the case of close large values of $D$ in Fig.~\ref{fig:STDP1}(d2).
The initial value of $M_s$ for $D=0.7$ is a little larger than that for $D=0.77$. But, as the time $t$ increases, $M_s$ for $D=0.7$ increases thanks to long-term potentiation, while $M_s$ for $D=0.77$ decreases due to long-term depression.
Furthermore, we note that $M_s$ for $D=0.7$ (0.25) increases (decreases) due to long-term potentiation (long-term depression), although their initial values of $M_s$ are nearly the same [see Fig.~\ref{fig:STDP1}(d3)]. This seems to
occur because MFRs of individual neurons for $D=0.7$ are higher than those for $D=0.25$. For the case of higher MFR, the distribution of $\{ \Delta t_{ij} \}$ may be narrower, which seems to
lead to long-term potentiation  \cite{STDP7,Tass2}. Figures \ref{fig:STDP1}(e1)-\ref{fig:STDP1}(e6) show the histograms for MFRs $f_i$ of individual neurons in black color for various values of $D$; the MFR
$f_i$ for each neuron is obtained through averaging of $10^5$ msec after the saturation time ($t=2000$ sec) and the bin size for each histogram is 0.2 Hz. For comparison, initial distributions for $\{ f_i \}$ are shown in gray
color. In the case of long-term potentiation ($D=0.27,$ 0.3, 0.5, and 0.7), both the population frequency $f_p$ of the IPSR kernel estimate $R(t)$ and the degree of SSS increase, in comparison with those in the absence of STDP
[compare Figs.~\ref{fig:STDP1}(f2)-\ref{fig:STDP1}(f5) and Figs.~\ref{fig:STDP1}(g2)-\ref{fig:STDP1}(g5) with Figs.~\ref{fig:NSTDP1}(b3)-\ref{fig:NSTDP1}(b6) and Figs.~\ref{fig:NSTDP1}(c3)-\ref{fig:NSTDP1}(c6)].
As a result, the population-averaged MFRs $\langle f_i \rangle$ become higher (i.e., black histograms lie on the right side of the initial gray histograms), and their standard deviations from $\langle f_i \rangle$
are generally smaller (i.e., widths of the black histograms are generally narrower) except for the case of small $D$ (=0.27) where the standard deviations are the same in both the presence and the absence of STDP.
For the case of long-term depression ($D=0.25$ and 0.77), the population states become desynchronized. Without any coherent synaptic inputs, individual neurons fire randomly mainly due to noise. Hence, for large $D$ (=0.77)
the population-averaged MFR $\langle f_i \rangle$ increases, while  $\langle f_i \rangle$ for small $D$ (=0.25) becomes smaller. For both cases, the standard deviations of their distributions become larger
because of the noise effect. Neural synchronization may be well visualized in the raster plot of spikes, and the corresponding IPSR kernel estimate $R(t)$ shows the population behaviors well.
Figures \ref{fig:STDP1}(f1)-\ref{fig:STDP1}(f6) and Figures \ref{fig:STDP1}(g1)-\ref{fig:STDP1}(g6) show raster plots of spikes and the corresponding IPSR kernel estimates $R(t)$ for various values of $D$,
respectively. When compared with Figs.~\ref{fig:NSTDP1}(b2)-\ref{fig:NSTDP1}(b7) and Figs.~\ref{fig:NSTDP1}(c2)-\ref{fig:NSTDP1}(c7) in the absence of STDP, the degrees of SSS for the case
of long-term potentiation ($D=0.27,$ 0.3, 0.5 and 0.7) are increased so much, while in the case of long-term depression ($D=0.25$ and 0.77) the population states become desynchronized.

\begin{figure}
\includegraphics[width=0.7\columnwidth]{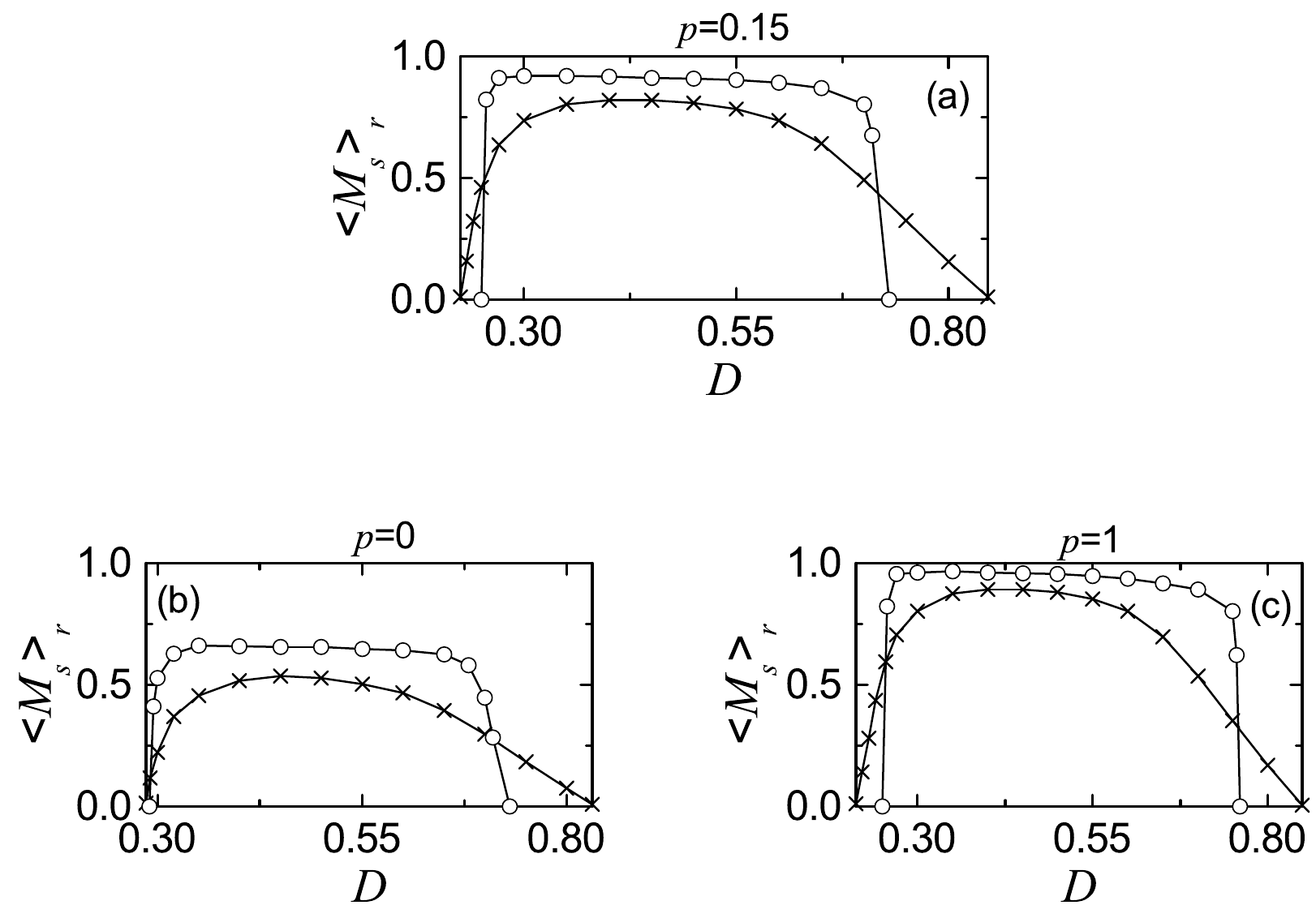}
\caption{Effects of the additive STDP on the statistical-mechanical spiking measure $M_s$ for various values of $p$.
Plots of the statistical-mechanical spiking measure $\langle M_s \rangle_r$ (represented by open circles) versus $D$ for $p$ = (a) 0.15, (b) 0, and (c) 1.
$M_s$ is obtained by following  $3 \times 10^3$ stripes in the raster plot of spikes after the saturation time 2000 sec in each realization. For  comparison,
$\langle M_s \rangle_r$ in the absence of STDP are shown in crosses.
}
\label{fig:STDP2}
\end{figure}

We characterize the SSS in terms of the statistical-mechanical spiking measure $M_s$, which is also compared with the case without STDP. Figure \ref{fig:STDP2}(a) shows the plot of $\langle M_s \rangle_r$
(represented by open circles) versus $D$ in the SWN with $p=0.15$; for  comparison, $\langle M_s \rangle_r$ in the absence of STDP are shown in crosses. A Matthew effect in synaptic plasticity occurs via
a positive feedback process. Good synchronization gets better through long-term potentiation, while bad synchronization gets worse through long-term depression. Consequently, a rapid step-like transition to SSS takes place, which is in contrast
to the relatively smooth transition in the absence of STDP. For comparison with the case of SWN with $p=0.15$, we also consider the cases of $p=0$ (regular lattice) and $p=1$ (random graph). The regular lattice is highly
clustered but large world, while the random graph is poorly clustered but small world. As a cluster-friendly extension of the random graph, the SWN has both high clustering and short path length.
Figures \ref{fig:STDP2}(b) and  \ref{fig:STDP2}(c) show plots of $\langle M_s \rangle_r$ (denoted by open circles) for $p=0$ and 1, respectively; $\langle M_s \rangle_r$ in the absence of STDP are also
shown in crosses. As in the SWN, Matthew effects in synaptic plasticity occur in both cases of $p=0$ and 1, and hence rapid transitions to SSS occur. The degree of SSS (given by $\langle M_s \rangle_r$)
for the case of SWN with $p=0.15$ is much larger than that for the case of $p=0$ because the average path length $L$ for the SWN with $p=0.15$ is much shorter than that for $p=0$. $L$ is dramatically decreased
with increasing $p$, and hence $L$ for $p=0.15$ is close to that for $p=1$. As a result, $\langle M_s \rangle_r$ for $p=0.15$ is close to that for $p=1$. Moreover, as $p$ is increased, transitions to SSS become
more rapid due to the increased Matthew effect.

\begin{figure}
\includegraphics[width=\columnwidth]{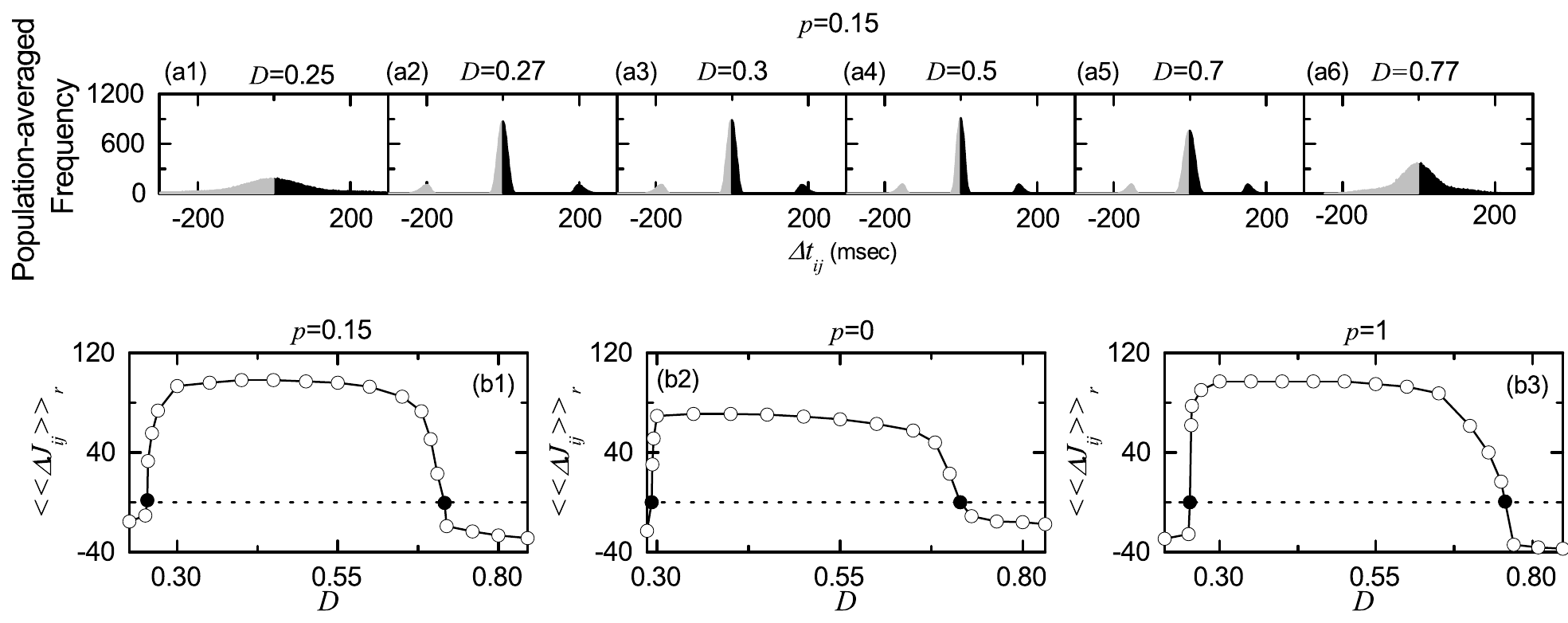}
\caption{Distributions of microscopic time delays between the pre- and the post-synaptic spike times and synaptic modifications. (a1)-(a6) Population-averaged histograms
for the distributions of time delays $\{ \Delta t_{ij} \}$ during the time interval from $t=0$ to the saturation time ($t=2000$ sec) for various values of $D$ in the SWN with
$p=0.15$; black and gray regions represent long-term potentiation and long-term depression, respectively. Plots of the population-averaged synaptic modifications $\langle \langle \Delta J_{ij} \rangle \rangle_r$
versus $D$ for $p=$ (b1) 0.15, (b2) 0, and (b3) 1. The values of $\langle \langle \Delta J_{ij} \rangle \rangle_r$ are obtained from the population-averaged histograms for distributions
of $\{ \Delta t_{ij} \}$, and solid circles denote the lower threshold $D^*_l$ and the higher threshold $D^*_h$.
}
\label{fig:MI1}
\end{figure}

From now on, we make an intensive investigation on emergences of long-term potentiation and long-term depression of synaptic strengths via microscopic studies based on the distributions of time delays $\{ \Delta t_{ij} \}$ between the pre- and
the post-synaptic spike times. Figures \ref{fig:MI1}(a1)-\ref{fig:MI1}(a6) show population-averaged histograms $H(\Delta t_{ij})$ for the distributions of time delays $\{ \Delta t_{ij} \}$ during the time
interval from $t=0$ to the saturation time ($t=2000$ sec) for various values of $D$ in the SWN with $p=0.15$: for each synaptic pair, its histogram for the distribution of $\{ \Delta t_{ij} \}$ is obtained, and then we get the
population-averaged histogram via averaging over all synaptic pairs. Here, black and gray regions represent long-term potentiation and long-term depression, respectively. In the case of long-term potentiation ($D=0.27,$ 0.3, 0.5, and 0.7), 3 peaks appear: one main central peak and two left and right minor peaks. When the pre- and the post-synaptic spike times appear in the same spiking stripe in the raster plot of spikes, its time delay $\Delta t_{ij}$ lies in the main peak; long-term potentiation and long-term depression may occur depending on the sign of $\Delta t_{ij}$. On the other hand, time delays $\Delta t_{ij}$ lie in the minor peaks when the pre- and the post-synaptic spike times appear in the different nearest-neighboring spiking stripes. If the pre-synaptic stripe precedes the post-synaptic stripe (causality), then its time delay $\Delta t_{ij}$ lies in the right minor peak (long-term potentiation); otherwise, it lies in the left minor peak (long-term depression). For the case of long-term depression ($D=0.25$ and 0.77), the population states become desynchronized due to overlap of spiking stripes in the raster plot of spikes. Consequently, the main peak in the histogram becomes merged with the left and the right minor peaks, and then only one broadened main peak appears, in contrast to the case of long-term potentiation. The population-averaged synaptic modification $\langle \langle \Delta J_{ij} \rangle \rangle_r$ [during the time interval from $t=0$ to the saturation time ($t=2000$ sec)] may be directly obtained from the above histogram $H(\Delta t_{ij})$:
\begin{equation}
  \langle \langle \Delta J_{ij} \rangle \rangle_r \simeq \sum_{\rm {bins}} H(\Delta t_{ij}) \cdot \Delta J_{ij} (\Delta t_{ij}).
\end{equation}
Figure \ref{fig:MI1}(b1) shows a plot of $\langle \langle \Delta J_{ij} \rangle \rangle_r$ [obtained from $H(\Delta t_{ij})$] versus $D$ for $p=0.15$; solid circles represent the lower and the higher
thresholds $\widetilde{D}_l$ and $\widetilde{D}_h$ for long-term potentiation and long-term depression (where $\langle \langle \Delta J_{ij} \rangle \rangle_r = 0$) which are the same as those in Fig.~\ref{fig:STDP1}(c). Then, population-averaged limit
values of synaptic strengths $\langle \langle J_{ij}^* \rangle \rangle_r$ are given by $J_0 + \delta~\langle \langle \Delta J_{ij} \rangle \rangle_r$, which agree well with the directly-obtained values in
Fig.~\ref{fig:STDP1}(c). Similarly, for $p=0$ (regular lattice) and 1 (randon graph), we also obtain population-averaged synaptic modification $\langle \langle \Delta J_{ij} \rangle \rangle_r$ from the population-averaged histograms $H(\Delta t_{ij})$ for the distributions of time delays $\{ \Delta t_{ij} \}$ during the time interval from $t=0$ to the saturation time ($t=2000$ sec), which are shown in Figs.~\ref{fig:MI1}(b2) and \ref{fig:MI1}(b3),
respectively. As $p$ is increased from 0, the range of long-term potentiation [i.e., ($\widetilde{D}_l$, $\widetilde{D}_h$)] becomes wider and most synaptic modifications for the long-term potentiation are also increased.

\begin{figure}
\includegraphics[width=\columnwidth]{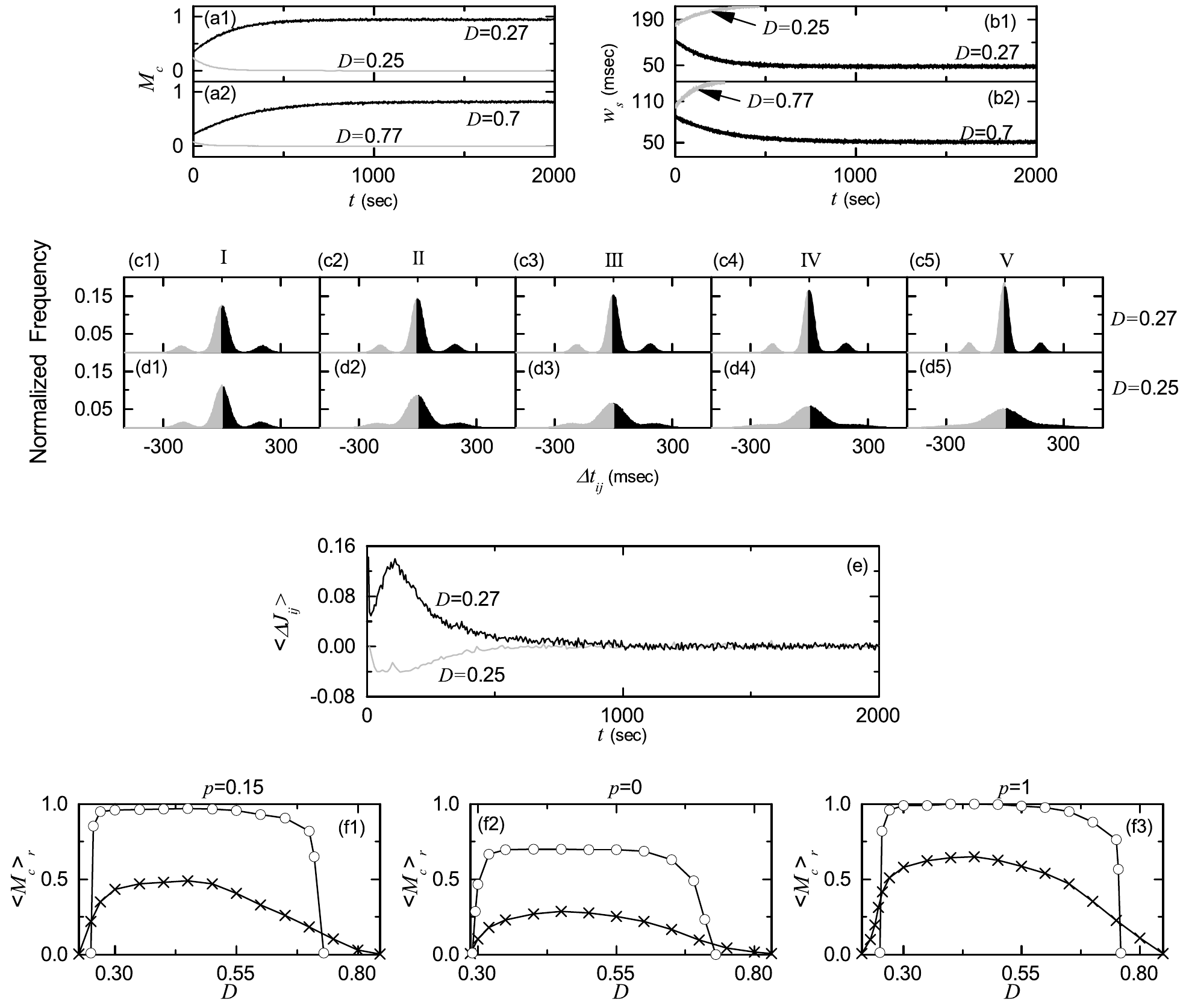}
\caption{Microscopic cross-correlations between synaptic pairs. Time-evolutions of the microscopic correlation measure $M_c(t)$ for (a1) $D = 0.25$ and 0.27 and (a2) $D = 0.7$ and 0.77
when $p = 0.15$. Time-evolutions of the width $w_s(t)$ of the spiking stripes in the raster plot of spikes for (b1) $D = 0.25$ and 0.27 and (b2) $D = 0.7$ and 0.77 in the SWN with
$p = 0.15$. Time-evolutions of the normalized histogram $H(\Delta t_{ij})$ for the distributions of time delays $\{ \Delta t_{ij} \}$ between the pre- and the post-synaptic spike times for $D = 0.27$ in (c1)-(c5)
and for $D = 0.25$ in (d1)-(d5) when $p = 0.15$; 5 stages are shown in I ($18 \sim 1078$ msec for $D=0.27$ and $12 \sim 1177$ msec for $D=0.25$), II ($100012 \sim 101032$ msec for $D=0.27$
and $100002 \sim 101107$ msec for $D=0.25$), III ($300046 \sim 301036$ msec for $D=0.27$ and $300022 \sim 301067$ msec for $D=0.25$), IV ($500012 \sim 500952$ msec for $D=0.27$ and
$500008 \sim 500993$ msec for $D=0.25$), and V ($1000032 \sim 1000962$ msec for $D=0.27$ and $1000002 \sim 1000927$ msec for $D=0.25$). (e) Time-evolutions of population-averaged synaptic modifications
$\langle\Delta J_{ij}(t) \rangle$ for $D = 0.27$ (black line) and for $D = 0.25$ (gray line) when $p = 0.15$. Plots of $\langle M_c \rangle_r$ (represented by open circles) versus $D$
in the saturated limit case for $p =$ (f1) 0.15, (f2) 0, and (f3) 1. For comparison, $\langle M_c \rangle_r$ in the absence of STDP are also shown in crosses.
}
\label{fig:MI2}
\end{figure}

Finally, we study the effect of STDP on the microscopic pair-correlation $C_{ij}(\tau)$ between the pre- and the post-synaptic IISRs (instantaneous individual spike rates) for the $(i,j)$ synaptic
pair. For obtaining dynamical pair-correlations, each spike train of the $i$th neuron is convoluted with a Gaussian kernel function $K_h(t)$ of band width $h$ to get a smooth estimate of IISR $r_i(t)$:
\begin{equation}
r_i(t) = \sum_{s=1}^{n_i} K_h (t-t_{s}^{(i)}),
\label{eq:IISR}
\end{equation}
where $t_{s}^{(i)}$ is the $s$th spiking time of the $i$th neuron, $n_i$ is the total number of spikes for the $i$th neuron, and $K_h(t)$ is given in Eq.~(\ref{eq:Gaussian}). Then, the normalized
temporal cross-correlation function $C_{ij}(\tau)$ between the IISR kernel estimates $r_i(t)$ and $r_j(t)$ of the $(i,j)$ synaptic pair is given by:
\begin{equation}
C_{ij}(\tau) = \frac{\overline{\Delta r_i(t+\tau) \Delta r_j(t)}}{\sqrt{\overline{\Delta r^2_i(t)}}\sqrt{\overline{\Delta r^2_j(t)}}},
\end{equation}
where $\Delta r_i(t) = r_i(t) - \overline{r_i(t)}$ and the overline denotes the time average.
Then, the microscopic correlation measure $M_c,$ representing the average ``in-phase'' degree between the pre- and the post-synaptic pairs, is given by the average value of $C_{ij}(0)$ at the zero-time
lag for all synaptic pairs:
\begin{equation}
M_c = \frac{1}{N_{syn}} \sum_{(i,j)} C_{ij}(0),
\label{eq:CM}
\end{equation}
where $N_{syn}$ is the total number of synapses.
Time-evolutions of the microscopic correlation measures $M_c(t)$ for the population states are shown in Figs.~\ref{fig:MI2}(a1)-\ref{fig:MI2}(a2) for the case of SWN with $p=0.15$. Data for calculation of $M_c(t)$ are
obtained via averages during successive 5 global cycles of the IPSR kernel estimate $R(t)$ for the case of long-term potentiation, while the data for the case of long-term depression are obtained through averages during successive 5 global cycles
of $R(t)$ for $t < 500$ sec and during successive 100 global cycles for $t > 500$ sec. For the case of close small values of $D$ in Fig.~\ref{fig:MI2}(a1), the initial value of $M_c$ for $D=0.27$ is
a little larger than that for $D=0.25$. However, with increasing time $t$, $M_c$ for $D=0.27$ increases, and it approaches a limit value. On the other hand, $M_c$ for $D=0.25$ decreases with time $t$, and
it seems to approach zero. A similar one occurs in the case of close large values of $D$ in Fig.~\ref{fig:MI2}(a2). The initial value of $M_c$ for $D=0.7$ is a little larger than that for $D=0.77$.
But, as the time $t$ increases, $M_c$ for $D=0.7$ increases, while $M_c$ for $D=0.77$ decreases. Enhancement (suppression) in $M_c$ results in increase (decrease) in the average in-phase degree
between the pre- and the post-synaptic pairs. Then, widths of spiking stripes in the raster plot of spikes decrease (increase) due to enhancement (suppression) of $M_c$.
Figures \ref{fig:MI2}(b1)-\ref{fig:MI2}(b2) show time-evolutions of the width $w_s(t)$ of the spiking stripes for $p=0.15$; $w_s(t)$ is obtained through averaging the widths
of spiking stripes during successive 5 global cycles of $R(t)$. For $D=0.27$ and 0.7, $w_s(t)$ decreases thanks to enhancement in $M_c$, which leads to narrowed distribution of time delays
$\{ \Delta t_{ij} \}$ between the pre- and the post-synaptic spike times. Consequently, long-term potentiation may occur. In contrast, for $D=0.25$ and 0.77, $w_s(t)$ increases due to suppression in $M_c$
(calculations of $w_s(t)$ for $D=$ 0.25 and 0.77 are made until $t \simeq$ 464 sec and 273 sec, respectively, when spiking stripes begin to overlap), which results in widened distribution of time delays
$\{ \Delta t_{ij} \}$. As a result, long-term depression may takes place.

Time-evolutions of normalized histograms $H(\Delta t_{ij})$ for the distributions of time delays $\{ \Delta t_{ij} \}$ are shown for $D=0.27$ in Figs.~\ref{fig:MI2}(c1)-\ref{fig:MI2}(c5)
and for $D=0.25$ in Figs.~\ref{fig:MI2}(d1)-\ref{fig:MI2}(d5) when $p=0.15$; the bin size in each histogram is 2 msec. Here, we consider 5 stages [represented by I (starting from $\sim 0$ sec), II (starting from
$\sim 100$ sec), III (starting from $\sim 300$ sec), IV (starting from $\sim 500$ sec), and  V (starting from $\sim 1000$ sec)]; for more details, refer to the caption of Fig.~\ref{fig:MI2}.
At each stage, we get distribution for $\{ \Delta t_{ij} \}$ for all synaptic pairs during the 5 global cycles (about 1 sec) of the IPSR kernel estimate $R(t)$ and obtain normalized histogram by dividing the
distribution with the total number of synapses (=20000). For $D=0.27$ (long-term potentiation), 3 peaks appear in each histogram; main central peak and two left and right minor peaks. With increasing the time $t$
(i.e., with increase in the level of stage), peaks become narrowed, and then they become sharper. Two minor peaks also approach the main peak a little because the population frequency $f_p$ of $R(t)$ increases with
the stage. Furthermore, as the stage is increased, the main peak becomes more and more symmetric, and hence the effect of long-term potentiation in the black part tends to cancel out nearly the effect of long-term depression in the
gray part at the stage V. For $D=0.25$ (long-term depression), with increasing the level of the stage, peaks become wider and the merging-tendency between the peaks is intensified. At the stages IV and V, only one broad central
peak seems to appear. For the stage V, the effect of long-term potentiation in the black part tends to cancel out nearly the effect of long-term depression in the gray part because the broad peak is nearly symmetric. From these normalized histograms
$H(\Delta t_{ij})$ [obtained via averages during successive 5 global cycles of $R(t)$], we also get the population-averaged synaptic modification $\langle \Delta J_{ij} \rangle$ [$\simeq \sum_{\rm{bins}} H(\Delta t_{ij})
\cdot \Delta J_{ij} (\Delta t_{ij})$]. Figure \ref{fig:MI2}(e) shows time-evolutions of $\langle \Delta J_{ij} \rangle$ for $D=0.27$ (black curve) and $D=0.25$ (gray curve) when $p=0.15$. $\langle \Delta J_{ij}
\rangle$ for $D=0.27$ is positive, while it is negative for $D=0.25$. For both cases, they converge toward nearly zero at the stage V $(t \sim 1000$ sec) because the normalized histograms become nearly symmetric.
Then, the time evolution of population-averaged synaptic strength $\langle J_{ij} \rangle$ is given by $\langle J_{ij} \rangle = J_0 + \delta \sum_k \langle \Delta J_{ij}(k)
\rangle,$ where $k$ represents the average for the $k$th 5 global cycles of $R(t)$ and $J_0$(initial average synaptic strength)= 0.2. Time-evolutions of $\langle J_{ij} \rangle$ (obtained in this way)
for $D=0.27$ and 0.25 agree well with those in Fig.~\ref{fig:STDP1}(a). As a result, long-term potentiation (long-term depression) occurs for $D=0.27$ (0.25).

Figure \ref{fig:MI2}(f1) shows plots of $\langle M_c \rangle_r$ versus $D$ in the presence (open circles) and the absence (crosses) of STDP for the case of SWN with $p=0.15$; for comparison, the cases of $p=0$
(regular lattice) and $p=1$ (random graph) are also shown in Figs.~\ref{fig:MI2}(f2) and \ref{fig:MI2}(f3), respectively. The number of data used for the calculation of each temporal cross-correlation
function $C_{ij}(\tau)$ [the values of $C_{ij}(0)$ at the zero time lag are used for calculation of $M_c$] is $2^{16}$ (=65536) after the saturation time ($t=2000$ sec) in each realization. Like the
case of $\langle M_s \rangle_r$ in Fig.~\ref{fig:STDP2}, a Matthew effect also occurs in $\langle M_c \rangle_r$: good pair-correlation gets better, while bad pair-correlation gets worse. Hence, a step-like
transition occurs, in contrast to the case without STDP. As $p$ is increased from 0, such transitions become more rapid due to the increased Matthew effect. Since the average path length $L$ for $p=0.15$ is much
smaller than that for $p=0$ and close to that for $p=1$, the values of $\langle M_c \rangle_r$ on the top plateau for $p=0.15$ are much larger than those for $p=0$, and they are so close to those for $p=1$.

\subsection{Effects of The Multiplicative STDP on The SSS}
\label{sec:MSTDP}

In this subsection, we study the effect of multiplicative STDP (which depends on states) on the SSS in comparison with the (above) additive case. The coupling strength for each $(i,j)$ synapse is updated
with a multiplicative nearest-spike pair-based STDP rule \cite{Tass1,Multi}:
\begin{equation}
J_{ij} \rightarrow J_{ij} + (J^*-J_{ij})~|\delta~ \Delta J_{ij}(\Delta t_{ij})|.
\label{eq:MSTDP}
\end{equation}
Here, $\delta$ $(=0.005)$ is the update rate, $\Delta J_{ij}$ is the synaptic modification depending on the relative time difference $\Delta t_{ij}$ $(=t_i^{(post)} - t_j^{(pre)})$
between the nearest spike times of the post-synaptic neuron $i$ and the pre-synaptic neuron $j$ [time window for $\Delta J_{ij}$ is given in Eq.~(\ref{eq:TW})], and
$J^*=$ $J_h~(J_l)$ for the long-term potentiation (long-term depression) [$J_h(=1.0)$ and $J_l(=0.0001)$ is the higher (lower) bound of $J_{ij}$ (i.e., $J_{ij} \in [J_l, J_h])$].
For this multiplicative case, the bounds for the synaptic strength $J_{ij}$ become soft, because a change in synaptic strengths scales linearly with the distance to the
higher and the lower bounds, in contrast to the hard bounds for the case of additive STDP (without dependence on states).

\begin{figure}
\includegraphics[width=\columnwidth]{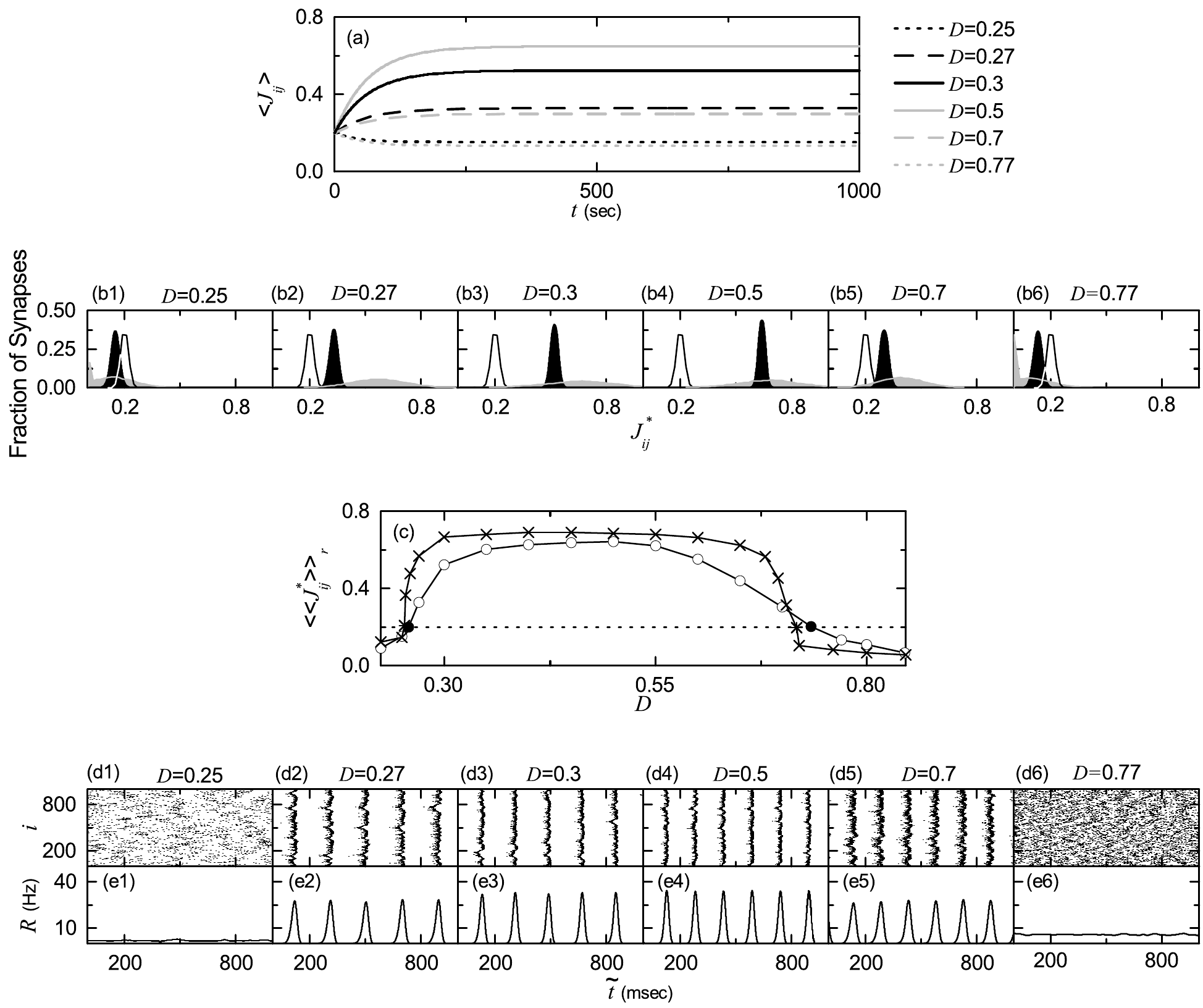}
\caption{Effects of the multiplicative STDP on the SSS in the SWN with $p=0.15$.
(a) Time-evolutions of population-averaged synaptic strengths $\langle J_{ij} \rangle$ for various values of $D$. (b1)-(b6) Histograms for the fraction of synapses versus
$J^*_{ij}$ (saturated limit values of $J_{ij}$ at $t = 500$ sec) for various values of $D$ (black region); for comparison, distributions of $\{ J^*_{ij} \}$ for the case of the additive STDP
and the initial distributions of $J_{ij}$ are also shown in gray regions and in black curves, respectively. (c) Plots of population-averaged limit values of synaptic strengths
$\langle \langle J_{ij}^* \rangle \rangle_r$ (denoted by open circles) versus $D$. The horizontal dotted line denotes the initial values of $\langle J_{ij} \rangle$ [$J_0~(= 0.2)$]. Solid circles represent threshold values
for the long-term potentiation and long-term depression of synaptic strengths (where $\langle \langle J_{ij}^* \rangle \rangle_r = J_0$). For comparison, $\langle \langle J_{ij} \rangle \rangle_r$ and threshold values for the long-term potentiation and long-term depression in the case of the additive STDP are also shown in crosses and stars, respectively. Raster plots of spikes in (d1)-(d6) and IPSR kernel estimates $R(t)$ in (e1)-(e6) for various values of $D$ after the saturation time,
where $t=t^*$ (saturation time=500 sec) + $\widetilde{t}$.
}
\label{fig:MULTI1}
\end{figure}

Figure \ref{fig:MULTI1}(a) shows time-evolutions of population-averaged synaptic strengths $\langle J_{ij} \rangle$ for various values of $D$ in the SWN with $p=0.15$. For $D=0.27,$ 0.3, 0.5 and 0.7,
$\langle J_{ij} \rangle$ increases above its initial value $J_0$ (= 0.2), and it approaches a saturated limit value $\langle J_{ij}^* \rangle$ nearly at $t=500$ sec. As a result, long-term potentiation occurs for these values of $D$.
On the other hand, for $D=0.25$ and 0.77 $\langle J_{ij} \rangle$ decreases below $J_0$, and approaches a saturated limit value $\langle J_{ij}^* \rangle$. Consequently, long-term depression occurs for these values of $D$.
When compared with the additive case in Fig.~\ref{fig:STDP1}(a), the saturation time is shorter and deviations of the saturated limit values $J_{ij}^*$ from $J_0$ are smaller due to the soft bounds.
Histograms for fraction of synapses versus $J_{ij}^*$ (saturated limit values of $J_{ij}$ at $t=500$ sec) for $p=0.15$ are shown in black regions for various values of $D$ in Figs.~\ref{fig:MULTI1}(b1)-\ref{fig:MULTI1}(b6);
the bin size for each histogram is 0.02. For comparison, distributions of $\{ J^*_{ij} \}$ for the case of the additive STDP and initial Gaussian distributions (mean $J_0$= 0.2 and standard deviation $\sigma_0$= 0.02)
of $\{ J_{ij} \}$ are also shown in gray regions and in black curves, respectively. As in the case of additive STDP, long-term potentiation occurs for $D=0.27,$ 0.3, 0.5, and 0.7, because their black histograms lie on the right side of the
initial black-curve histograms. However, these black histograms lie on the left side of the gray histograms for the case of additive STDP, and they are much narrower
than those for the additive case. Consequently, their population-averaged values $\langle J_{ij}^* \rangle$ and standard deviations $\sigma$ are smaller than those for the additive case,
because their variations in $J_{ij}$ are restricted due to soft bounds in comparison with hard bounds for the the additive case. Particularly, the standard deviations $\sigma$ for the multiplicative case are even smaller
than the initial ones $\sigma_0$ (= 0.02). On the other hand, for $D=0.25$ and 0.77 long-term depression occurs because the black histograms are shifted to the left side of the initial black-curve histograms. But, these black histograms
lie on the right side of the gray histograms for the case of additive STDP, and they are much narrower than those for the additive case. As a result, their population-averaged values
$\langle J_{ij}^* \rangle$  are larger than those for the additive case, due to soft bounds. Like the case of long-term potentiation, their standard deviations $\sigma$ are much smaller than those for the additive case and even smaller
than the initial ones $\sigma_0$ (= 0.02). Figure \ref{fig:MULTI1}(c) shows a plot of population-averaged limit values  $\langle \langle J_{ij}^* \rangle \rangle_r$ (denoted by open circles) of synaptic strengths versus $D$.
Here, the horizontal dotted line represents the initial average value of coupling strengths $J_0$ (= 0.2), and the lower and the higher thresholds $\widetilde{D}_l^*$ $(\simeq 0.258)$ and $\widetilde{D}_h^*$
$(\simeq 0.735)$ for long-term potentiation and long-term depression (where $\langle \langle J_{ij}^* \rangle \rangle_r = J_0$) are denoted by solid circles. Hence, long-term potentiation occurs in the range of ($\widetilde{D}_l^*$, $\widetilde{D}_h^*$); otherwise, long-term depression appears.
For comparison, the values of $\langle \langle J_{ij}^* \rangle \rangle_r$ for the additive case are also represented by crosses, and their lower and higher thresholds $\widetilde{D}_l$ $(\simeq 0.253)$ and $\widetilde{D}_h$ $(\simeq 0.717)$ are denoted by stars. When passing $\widetilde{D}_l^*$, a transition to long-term potentiation occurs for the multiplicative case, and then $\langle \langle J_{ij}^* \rangle \rangle_r$ increases in a relatively gradual way, in
comparison with the rapid (step-like) transition for the additive case. In the top region, a plateau (whose width is smaller than that for the additive case) appears,
then $\langle \langle J_{ij}^* \rangle \rangle_r$ decreases slowly (particularly, much slowly near the higher threshold $\widetilde{D}_l^*$ when compared with the additive case),
and a transition to long-term depression occurs as $\widetilde{D}_h^*$ is passed. Due to this gradual transition, $\widetilde{D}_l^*$ for the multiplicative case is a little larger than $\widetilde{D}_l$ for the additive case, and $\widetilde{D}_h^*$ is also relatively larger than $\widetilde{D}_h$. Hence, long-term potentiation for the multiplicative case occurs in a relatively wider range in comparison with the additive case, and most values of
$\langle \langle J_{ij}^* \rangle \rangle_r$ in the case of long-term potentiation are smaller than those for the additive case, due to soft bounds.

The effects of long-term potentiation and long-term depression on the SSS may be well visualized in the raster plot of spikes. Figures \ref{fig:MULTI1}(d1)-\ref{fig:MULTI1}(d6) and Figures \ref{fig:MULTI1}(e1)-\ref{fig:MULTI1}(e6) show raster plots
of spikes and the corresponding IPSR kernel estimates $R(t)$ for various values of $D$, respectively, in the case of $p=0.15$. When compared with Figs.~\ref{fig:NSTDP1}(b2)-\ref{fig:NSTDP1}(b7) and Figs.~\ref{fig:NSTDP1}(c2)-\ref{fig:NSTDP1}(c7) in the absence of STDP, as in the additive case, the degrees of SSS for the case of long-term potentiation ($D=0.27,$ 0.3, 0.5 and 0.7) are increased so much, while in the case of long-term depression ($D=0.25$ and 0.77) the population states become desynchronized. For the case of long-term potentiation, we also make comparison with additive case shown in Figs.~\ref{fig:STDP1}(f2)-\ref{fig:STDP1}(f5) and Figs.~\ref{fig:STDP1}(g2)-\ref{fig:STDP1}(g5).
For small $D$ (= 0.27), the value of $\langle \langle J_{ij}^* \rangle \rangle_r$ for the multiplicative case is much smaller than that for the additive case. Smaller $\langle \langle J_{ij}^* \rangle \rangle_r$
decreases the degree of synchronization. Hence, the widths of spiking stripes for the multiplicative case become a little wider than those for the additive case. However, for intermediate values of $D$ (= 0.3 and 0.5),
the standard deviations $\sigma$ for the distributions of $\{ J_{ij}^* \}$ in the multiplicative case are much smaller than those for the additive case, although their population-averaged values
$\langle \langle J_{ij}^* \rangle \rangle_r$ are still smaller. Effect of smaller standard deviation $\sigma$ (increasing the synchronization degree) balances out nearly the effect of smaller $\langle \langle J_{ij}^* \rangle \rangle_r$ (decreasing the degree of synchronization). Hence, the widths of spiking stripes become close to those for the additive case, which results in nearly the same degrees of SSS for both the multiplicative and the additive cases. For large $D$ (= 0.7), the widths of spiking stripes for the multiplicative case seem to be a little wider than those for the additive case due to its smaller population-averaged value $\langle \langle J_{ij}^* \rangle \rangle_r$. However, thanks to much smaller standard deviation $\sigma$ for the distribution of $\{ J_{ij}^* \},$ no scattered spikes appear between the spiking stripes for the multiplicative case, in contrast to the additive case. As a result, for the case of $D=0.7$, the whole degree of SSS in the multiplicative case seems to be a little higher than that for the additive
case, because the amplitude of the IPSR kernel estimate $R(t)$ is a little larger for the multiplicative case.

\begin{figure}
\includegraphics[width=\columnwidth]{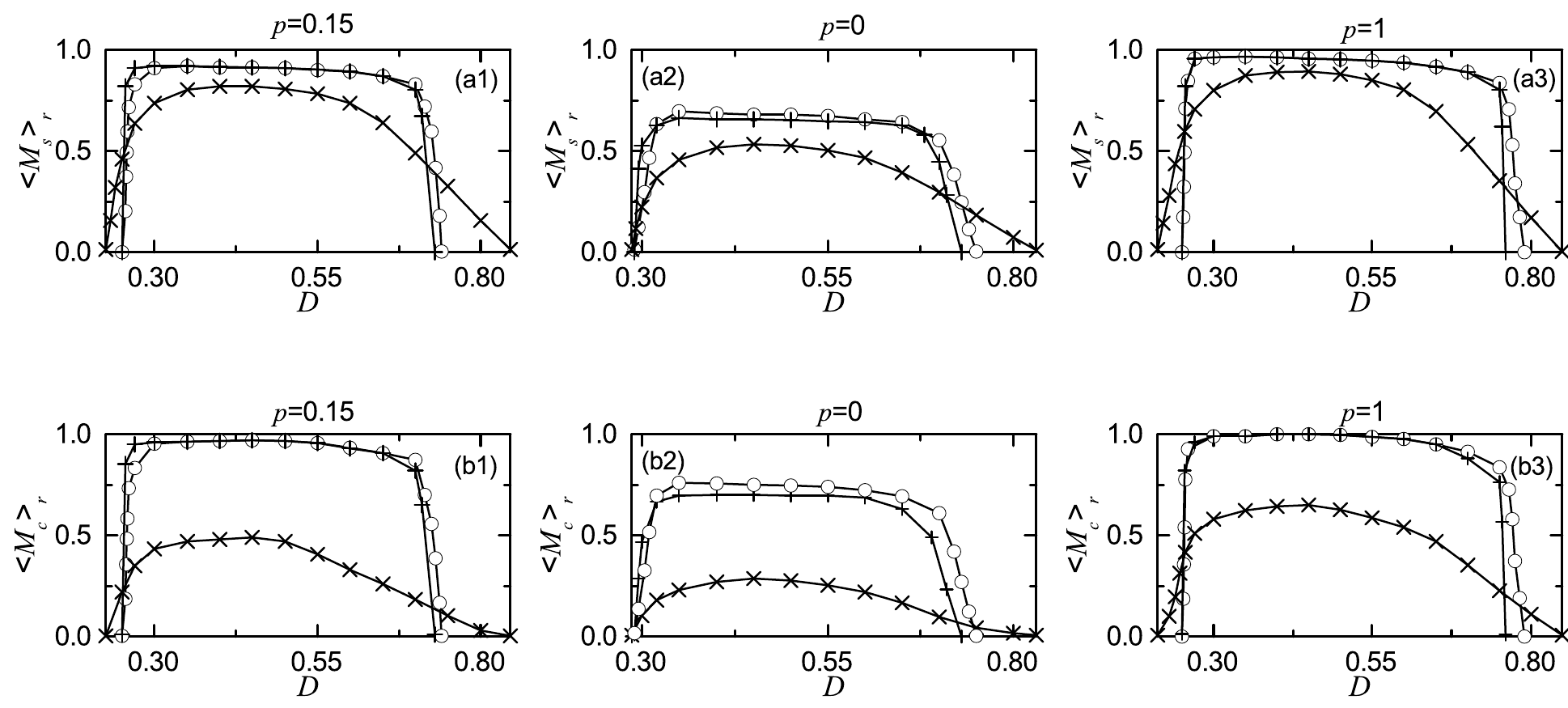}
\caption{Effects of the multiplicative STDP on the statistical-mechanical spiking measure $M_s$ and the microscopic correlation measure $M_c$ for various values of $p$. Plots of $\langle M_s \rangle_r$
(represented by open circles) versus $D$ for $p=$ (a1) 0.15, (a2) 0, and (a3) 1. For comparison, $\langle M_s \rangle_r$ in the absence of STDP and for the additive STDP case are also shown
in crosses and pluses, respectively. Plots of $\langle M_c \rangle_r$ (denoted by open circles) versus $D$ for $p=$ (b1) 0.15, (b2) 0, and (b3) 1. For comparison, $\langle M_c \rangle_r$ in
the absence of STDP and for the additive STDP case are shown in crosses and pluses, respectively.
}
\label{fig:MULTI2}
\end{figure}

Finally, we study the effects of multiplicative STDP on the statistical-mechanical spiking measure $M_s$ of Eq.~(\ref{eq:SM}) and the microscopic correlation measure $M_c$ of Eq.~(\ref{eq:CM}).
Figure \ref{fig:MULTI2}(a1) shows plots of $\langle M_s \rangle_r$ (represented by open circles for the multiplicative case) versus $D$ for $p=0.15$. For comparison, the values of $\langle M_s \rangle_r$ for the additive case and the case without STDP are also denoted by pluses and crosses, respectively. Here, $M_s$ is obtained by following $3 \times 10^3$ stripes in the raster plot of spikes after the saturation time ($t=$ 500 sec) in each realization. As in the case of additive STDP, a Matthew effect in synaptic plasticity occurs through a positive feedback process. Good synchronization gets better via long-term potentiation, while bad synchronization gets worse via long-term depression. As a result, a rapid
transition to SSS occurs, in contrast to the relatively smooth transition in the absence of STDP. However, changes near both ends are a little less rapid than those for the additive case, due to effects of soft bounds;
particularly, this type of change may be seen well near the right end. In most region of the top plateau in Fig.~\ref{fig:MULTI2}(a1), thanks to the effect of soft bounds, the standard deviations $\sigma$ for the distribution of $\{ J_{ij}^* \}$ in the multiplicative case are much smaller than those for the additive case, although their population-averaged values $\langle \langle J_{ij}^* \rangle \rangle_r$ are also smaller. Smaller standard deviation $\sigma$ (smaller $\langle \langle J_{ij}^* \rangle \rangle_r$) may increase (decrease) the degree of SSS. For most cases of long-term potentiation, these two effect are nearly balanced out, and hence the values of $\langle M_s \rangle_r$ are nearly the same for both the multiplicative and the additive cases. For comparison with the case of $p=0.15$ (SWN), we also consider the cases of $p=0$ (regular lattice with high clustering) and $p=1$ (random graph with short path length). As a cluster-friendly extension of the random graph, the SWN with $p=0.15$ has both high clustering and short path length. Figures \ref{fig:MULTI2}(a2) and  \ref{fig:MULTI2}(a3) show plots of $\langle M_s \rangle_r$
(denoted by open circles for the multiplicative case) for $p=0$ and 1, respectively; $\langle M_s \rangle_r$ for the additive case and in the absence of STDP are also shown in pluses and crosses, respectively. As in the case of $p=0.15$, Matthew effects in synaptic plasticity occur in both cases of $p=0$ and 1, and hence rapid transitions to SSS take place. Furthermore, with increasing $p$, transitions to SSS become more rapid due to the increased
Matthew effect. Most values of $\langle M_s \rangle_r$ (i.e., the degree of SSS) on the top plateau for the case of SWN ($p=0.15$) are much larger than those for the case of regular lattice ($p=0$) because of short path length $L$ for the SWN. These values of $\langle M_s \rangle_r$ are also close to those for the random graph ($p=1$) because $L$ for $p=0.15$ is close to that for the random graph.
Like the case of $p=0.15$, in most region of the top plateau for the case of $p=1$, the effects, associated with smaller population-averaged values $\langle \langle J_{ij}^* \rangle \rangle_r$ and smaller standard deviations
$\sigma$ for the distribution of $\{ J_{ij}^* \}$, are nearly balanced out, and hence the values of $\langle M_s \rangle_r$ are nearly the same for both the multiplicative and the additive cases. On the other hand, for the case of $p=0$, the values of $\langle M_s \rangle_r$ for the multiplicative case are a little larger than those for the additive case, because the effect, associated with the smaller standard deviations $\sigma$ (increasing the synchronization degree), outweights a little the effect, related to the smaller population-averaged values $\langle \langle J_{ij}^* \rangle \rangle_r$ (decreasing the degree of synchronization). Due to high clustering, zigzag patterns, intermingled with inclined partial stripes, appear in the raster plot of spikes for $p=0$. In the presence of high zigzagness, smaller standard deviations $\sigma$ for the multiplicative case seem to be more effective for reducing the degree of zigzagness, rather than larger population-averaged value $\langle \langle J_{ij}^* \rangle \rangle_r$ for the additive case.

Figure \ref{fig:MULTI2}(b1) shows plots of the microscopic correlation measure $\langle M_c \rangle_r$ for $p=0.15$ (SWN) in the multiplicative (``open circles'') and the additive (``pluses'') cases and in the
absence of STDP (``crosses''). For comparison, the cases of $p=0$ (regular lattice) and $p=1$ (random graph) are also shown in Figs.~\ref{fig:MULTI2}(b2) and \ref{fig:MULTI2}(b3), respectively. The number of data
used for the calculation of each temporal cross-correlation function $C_{ij}(\tau)$ [the values of $C_{ij}(0)$ at the zero time lag are used for calculation of $M_c$] is $2^{16}$ (=65536) after the saturation time
$t^*$ (=500 sec) in each realization. Like the case of $\langle M_s \rangle_r$, Matthew effects also occur in $\langle M_c \rangle_r$ for $p=0.15,$ 0, and 1: good pair-correlation gets better, while bad pair-correlation
gets worse. Hence, a rapid transition occurs, in contrast to the case without STDP. With increasing $p$, such transitions become more rapid due to the increased Matthew effect. Since the average path length $L$ for
$p=0.15$ is much smaller than $L$ for $p=0$ and close to $L$ for $p=1$, most values of $\langle M_c \rangle_r$ on the top plateau for $p=0.15$ are much larger than those for $p=0$, and they are so close to those for $p=1$.
Like the case of $\langle M_s \rangle_r$, in most region of the top plateau in Figs.~\ref{fig:MULTI2}(b1) and \ref{fig:MULTI2}(b3) for $p=0.15$ and 1, the effects, associated with smaller standard deviations $\sigma$ and
smaller population-averaged values $\langle J_{ij}^* \rangle$, are nearly balanced out, and hence most values of $\langle M_c \rangle_r$ on the top plateau are nearly the same for both the multiplicative and the additive cases.
On the other hand, for the case of $p=0$ with high local clustering, smaller standard deviations $\sigma$ are more effective for decreasing the zigzagness degree, and hence most values of $\langle M_c \rangle_r$ on the top plateau are a little larger for the multiplicative case.

\section{Summary}
\label{sec:SUM}
We considered an excitatory Watt-Strogatz SWN of subthreshold Izhikevich regular spiking neurons. Noise-induced firing patterns of subthreshold neurons may be used for encoding environmental stimuli. 
In previous works on the SSS (i.e., population synchronization between noise-induced spikings), synaptic strengths were static (i.e., synaptic plasticity was not considered). In contrast, adaptive dynamics of synaptic strengths in the present work are governed by the STDP. The effects of additive STDP (independent of states) on the SSS have been investigated in the SWN with $p=0.15$ by varying the noise intensity $D$. A Matthew effect in synaptic plasticity has been found to occur due to a positive feedback process. Good synchronization (with higher spiking measure $M_s$) gets better via long-term potentiation of synaptic strengths, while bad synchronization (with lower $M_s$) gets worse via long-term depression. Consequently, a step-like rapid transition to SSS occurs by changing $D$, in contrast to the relatively smooth transition in the absence of STDP.

Emergences of long-term potentiation and long-term depression of synaptic strengths were intensively investigated for the case of $p=0.15$ via microscopic studies based on both the distributions of time delays $\{ \Delta t_{ij} \}$ between the pre- and
the post-synaptic spike times and the pair-correlations between the pre- and the post-synaptic IISRs. For the case of long-term potentiation, three (separate) peaks (a main central peak and two left and right minor peaks) exist in the population-averaged histograms for the distributions of $\{ \Delta t_{ij} \}$, while a broad central peak appears via merging of the three peaks in the case of long-term depression. Then,
population-averaged synaptic modifications $\langle \Delta J_{ij} \rangle$ may be obtained from the population-averaged histograms, and they have been found to agree well with directly-calculated $\langle \Delta J_{ij} \rangle$. As a result, one may understand clearly how microscopic distributions of $\{ \Delta t_{ij} \}$ contribute to $\langle \Delta J_{ij} \rangle$. In addition, we are concerned about the microscopic correlation measure $M_c$, representing the in-phase degree between the pre- and the post-synaptic neurons, which are obtained from the pair correlations between the pre- and the post-synaptic IISRs. Like $M_s$, $M_c$ also exhibits a rapid transition due to a Matthew effect in the synaptic plasticity. Enhancement (suppression) of $M_c$ is directly related to decrease (increase) in the widths $w_s$ of spiking stripes in the raster plot of spikes. Then, distributions of $\{ \Delta t_{ij} \}$ become narrow (wide), which may lead to emergence of long-term potentiation (long-term depression). In this way, microscopic correlations between synaptic pairs are directly associated with appearance of long-term potentiation and long-term depression.

Effects of multiplicative STDP (which depends on states) on the SSS in the case of $p=0.15$ were also investigated in comparison with the additive case (independent of states). In this multiplicative case, the boundaries for the synaptic strength $J_{ij}$ become soft: a change in synaptic strengths scales linearly with the distance to the higher and the lower bounds, in contrast to the hard bounds for the additive case. Due to soft bounds, a gradual transition to long-term potentiation and long-term depression occurs, in comparison to the rapid transition for the additive case. Furthermore, thanks to the soft bounds, the standard deviations $\sigma$ for the distributions of saturated limit synaptic strengths $\{ J_{ij}^* \}$ are much smaller than those for the additive case. As a result of the smaller standard deviations $\sigma$ (increasing $M_s$), the degrees of SSS (given by $M_s$) for most cases of long-term potentiation become nearly the same as those in the additive case, although their population-averaged values $\langle J_{ij}^* \rangle$ are smaller. As in the case of $M_s$, a Matthew effect has also been found to occur in the microscopic correlation measure $M_c$. Good pair-correlation (with higher $M_c$) gets better via long-term potentiation, while bad synchronization (with lower $M_c$) gets worse via long-term depression. 

The results on $M_s$ and $M_c$ in the SWN with $p=0.15$ were also compared with those for $p=0$ (regular lattice) and $p=1$ (random graph). 
As in the case of $p=0.15$, Matthew effects also occur in both $\langle M_s \rangle_r$ and $\langle M_c \rangle_r$ for both cases of $p=0$ and 1. 
As a result, a rapid transition occurs, in contrast to the case without STDP. As $p$ is increased, such transitions become more rapid due to the increased Matthew effect. 
The average path length $L$ for $p=0.15$ is much smaller than $L$ for $p=0$ and close to $L$ for $p=1$. Hence, most values of $\langle M_s \rangle_r$ and $\langle M_c \rangle_r$ in the case of long-term potentiation        
for $p=0.15$ are much larger than those for $p=0$, and they are so close to those for $p=1$.

To get better insights on the results obtained via our numerical works, analytical works seem to be necessary. However, such analytical work is beyond the scope of present work, 
and it is left as a future research work. 

\section*{Acknowledgments}
This research was supported by the Basic Science Research Program through the National Research Foundation of Korea (NRF) funded by the Ministry of Education
(Grant No. 20162007688).

\end{document}